\documentstyle[12pt]{article} 

\newcommand{\be}{\begin{equation}}
\newcommand{\ee}{\end{equation}}
\newcommand{\bea}{\begin{eqnarray}}
\newcommand{\eea}{\end{eqnarray}}
\renewcommand{\theequation}{\arabic{section}.\arabic{equation}}
\newcommand{\lbl}[1]{\label{eq:#1}}
\newcommand{\rf}[1]{(\ref{eq:#1})}
\newcommand{\VAP}{\langle V \! A P\rangle}
\newcommand{\VVP}{\langle V V \! P\rangle}
\newcommand{\AAP}{\langle A A P\rangle}
\newcommand{\XPT}{ChPT}
\newcommand{\lag}{{\cal L}}
\newcommand{\order}{{\cal O}}
\newcommand{\F}{{\cal F}}
\newcommand{\G}{{\cal G}}
\renewcommand{\H}{{\cal H}}

\newcommand{\eps}{\epsilon}

\newcommand{\hv}{h}
\newcommand{\hp}{{\hat h}}

\newcommand{\qs}{q \!\!\! /}

\newcommand{\qbarq}{\langle{\overline\psi}\psi\rangle_0}

\def\thefootnote{\fnsymbol{footnote}}

\begin{document}
\begin{flushright}
August 21, 2001 \\  
hep-ph/0106034  
\end{flushright}

\vspace*{0.5cm}
\begin{center}
{\Large {\bf Resonance estimates of ${\cal O}(\mbox{\boldmath{$p^6$}})$
low-energy constants and QCD short-distance constraints}}\\[0.6cm]
Marc Knecht\footnote{knecht@cpt.univ-mrs.fr} and Andreas
Nyf\/feler\footnote{nyf\/feler@cpt.univ-mrs.fr} \\[0.3cm] 
Centre de Physique Th\'{e}orique, CNRS-Luminy, Case 907\\ 
    F-13288 Marseille Cedex 9, France
\begin{abstract}
Starting from the study of the low-energy and high-energy behaviours
of the QCD three-point functions $\VAP$, $\VVP$ and $\AAP$, several
${\cal O}({p}^6)$ low-energy constants of the chiral Lagrangian are
evaluated within the framework of the lowest meson dominance (LMD)
approximation to the large-$N_C$ limit of QCD. In certain cases,
values that differ substantially from estimates based on a resonance
Lagrangian are obtained. It is pointed out that the differences arise
through the fact that QCD short-distance constraints are in general
not correctly taken into account in the approaches using resonance
Lagrangians. We discuss the implications of our results for the 
${\cal O}(p^6)$ counterterm contributions to the vector form factor of
the pion and to the decay $\pi\to e \nu_e \gamma$, and for the
pion-photon-photon transition form factor. 
\end{abstract}

\end{center}

\renewcommand{\thefootnote}{\arabic{footnote}}
\setcounter{footnote}{0}


\section{Introduction}
 
In the chiral limit, the lightest pseudoscalar states of the hadronic
spectrum become the octet of massless Goldstone bosons resulting from
the spontaneous breaking of the chiral $SU(3)_L\times SU(3)_R$ global
symmetry of the QCD Lagrangian towards its diagonal subgroup $SU(3)_V$
of vector symmetries. This well-known fact \cite{tHooft,Vafa} allows
to describe the interactions of the light pseudoscalar mesons at low
energies in terms of an effective Lagrangian
\cite{EffLag,GL_84_85}. The latter involves the pseudoscalar fields,
described by a unitary matrix $U(x)$, and transforming under a
nonlinear representation of the chiral symmetry group, as well as the
sources, $v_{\mu}(x)$, $a_{\mu}(x)$, of the light-quark vector and
axial currents, and $s(x)$, $p(x)$, of the scalar and pseudoscalar
densities of QCD \cite{GL_84_85}.  Matrix elements of these currents
between pseudoscalar states, or scattering amplitudes involving these
light states only, can be computed in a systematic way in the
low-energy theory as long as all momentum transfers $p^2$ are
sufficiently small, $p^2\ll\Lambda_H^2$, where $\Lambda_H\sim 1$ GeV
is the typical scale at which the non-Goldstone bound states of QCD
are formed.  Since the (running) light quark masses $m_q(\mu)$ are also
small as compared to this scale, $m_{u,d,s}(\Lambda_H)\ll\Lambda_H$,
the effective Lagrangian ${\cal L}^{\mbox{\scriptsize eff}}$ can be
organized as an expansion in powers of derivatives of the field $U(x)$
and powers of the light quark masses,
\bea 
{\cal L}^{\mbox{\scriptsize eff}}\,=\,\sum_{k,l}{\cal L}_{(k,l)}\,,
\nonumber 
\eea
with
\bea \label{lag_kl} 
{\cal L}_{(k,l)}\,\sim\,\bigg(\frac{p}{\Lambda_H}\bigg)^k
\bigg(\frac{m_q}{\Lambda_H}\bigg)^l\,. 
\eea 
The presently available studies on the structure of the low-energy
effective Lagrangian involve, beyond the lowest order terms, the
pieces ${\cal L}_{(4,0)}$, ${\cal L}_{(2,1)}$ and ${\cal L}_{(0,2)}$
\cite{GL_84_85}, ${\cal L}_{(6,0)}$ and ${\cal
L}_{(4,1)}$~\cite{Fearing_Scherer,ChPT_p6}, ${\cal L}_{(2,2)}$ and
${\cal L}_{(0,3)}$~\cite{Fuchs:1991cq,Fearing_Scherer,ChPT_p6}, as
well as ${\cal L}_{(0,4)}$~\cite{Fuchs:2000hq}.  The structure of each
${\cal L}_{(k,l)}$ is entirely fixed by the chiral symmetry properties
of QCD \cite{Leutwyler94}, but involves coefficients, the so called
low-energy constants,~\footnote{Some of the counterterms involving
the external sources only and no  pseudoscalar fields actually
rather correspond to ``high-energy constants'', since they describe
the (perturbative) short-distance ambiguities of some QCD
correlators. At $\order(p^4)$, this concerns the constants $H_1$ and
$H_2$ of Ref.~\cite{GL_84_85}. We exclude this type of counterterms 
from the present
discussion.} which are not determined from symmetry requirements
alone. The predictive power of the effective theory therefore hinges
to quite some extent on the knowledge of these low-energy
constants. At ${\cal O}(p^4)$, the values of most of the low-energy
constants were extracted from data. The proliferation of low-energy
constants at ${\cal O}(p^6)$ makes such an approach unrealistic.

On the other hand, it is a general property that these low-energy
constants correspond to the coefficients of the Taylor expansion, with
respect to the momenta, of some QCD correlation functions, once the
singularities (poles and discontinuities) associated with the
contributions of low-momentum pseudoscalar intermediate states have
been subtracted.  The characteristic feature of the Green's functions
that are actually involved is that they are {\it order parameters} of
the spontaneous breaking of chiral symmetry. Thus, they do not receive
contributions from perturbative QCD at large momentum transfers, but
rather exhibit a smooth behaviour at short distances. The low-energy
constants are thus expected to be sensitive to the physics in the
intermediate energy region, that is to the spectrum of mesonic
resonances in the mass region around the hadronic scale $\Lambda_H$.
This basic observation underlies, in some way or another, most
attempts to estimate the values of the low-energy constants from
resonance data (for an introduction to the vast bibliography on this
subject, we refer the reader to the review articles
\cite{Meissner:1993ah,Ecker:1995gg,Pich1998}). 

It has become customary to describe the effects of resonance states
within a Lagrangian framework, by introducing, besides the Goldstone
boson field $U(x)$, additional local fields associated with the meson
states.  While there exists a systematic way \cite{CCWZ} to construct
fields which have the appropriate transformation properties under
chiral transformations and invariant Lagrangians, there is however no
restriction from chiral symmetry as to the number of fields and the
order of derivatives thereof involved in the terms which describe the
interaction among resonances or between the resonances and the
Goldstone bosons. In addition, the construction of Ref. \cite{CCWZ}
leaves open the question of the choice of the Lorentz group
representation for the field describing a meson state of a given spin.
This lack of restrictions has led to many proposals concerning a
Lagrangian description of interacting Goldstone bosons and mesons (the
literature on this subject can be traced back from several reviews
\cite{Gasio69,Meissner:1988ge,Bando88} and from the articles
\cite{Donoghue:1989ed,Ecker:1989te,Birse:1996hd}).  It has been
pointed out in Ref. \cite{Ecker:1989yg} that restrictions actually can
be introduced, {\it via} the QCD {\it short-distance} properties of
the relevant Green's functions or form factors, and by requiring that
these properties are satisfied by the same objects constructed with
the help of the Lagrangian involving resonances. This aspect has been
taken into account in a rather systematic manner at the level of the
${\cal O}(p^4)$ counterterms~\cite{Ecker:1989yg}. Although resonance
estimates have been given for several ${\cal O}(p^6)$ low-energy
constants, the importance of implementing the appropriate QCD
short-distance constraints has not always been stressed. The purpose
of the present study is precisely to address this aspect.

Although quite practical and useful, since it guarantees that some
general properties (locality, analyticity) of quantum field theories
are correctly taken into account, a Lagrangian formalism is not
absolutely necessary in order to estimate the contributions of the QCD
resonance states to the low-energy constants. In this article, we
shall in fact consider a different approach, working directly with the
appropriate Green's functions. We shall however retain the general
features and assumptions that underlie the Lagrangian approach, and
that we briefly recall. First, one usually considers zero-width
resonances, and takes into account the contributions from
one-resonance states, which produce only poles in the corresponding
correlation functions. There exists a well-defined framework where
this kind of restrictions arises in a natural way~\cite{witten},
namely the large-$N_C$ limit of QCD \cite{tHooftNc}. On the other
hand, working in this limit still requires to consider, in each
channel, an infinite number of resonances, with masses and coupling
constants adjusted such as to reproduce the QCD perturbative continuum
at high momentum transfers. Of course, such a point of view is rather
ambitious, since it amounts to solving QCD in the large-$N_C$ limit, a
reputedly difficult task. We shall adopt a more modest attitude,
assuming that in each channel a few lowest-lying resonances give
already the main contribution (approaches involving an infinite number
of zero-width resonances, with various additional assumptions about
their spacing, decay constants, etc., can be found in
Refs.~\cite{Geshkenbein,Bolokhov,Golterman:2001nk}). The number of
resonances to be considered in each channel will be taken as the
minimal (finite) number necessary to satisfy the requirements set by,
say, the leading QCD short-distance constraints for the Green's
functions under consideration, and possibly other constraints that one
may wish to impose. This {\it minimal hadronic ansatz} (MHA)
approximation may be well justified in the case of Green's functions
which are order parameters, free of perturbative contributions (for
Green's functions which are not order parameters, the QCD continuum
contribution has to be included as well, see
Refs.~\cite{Peris:1998nj,Golterman:2000au}). In fact, in many cases,
this minimal hadronic ansatz can be reduced to retaining, in each
channel, a single resonance. At the ${\cal O}(p^4)$ level, this {\it
lowest meson dominance} (LMD) approximation to large-$N_C$ QCD has
been tested successfully in several instances
\cite{Donoghue:1989ed,Ecker:1989te,Ecker:1989yg,Peris:1998nj}. As we
shall see in the examples treated in the present work, depending on
the constraints one wishes to implement on the Green's functions under
consideration, this simplest LMD approximation may however not always
be sufficient. A second feature common to the Lagrangian and to the
LMD or MHA approximations is the fact that the estimates of the
low-energy constants do not reproduce their scale dependence. The
latter, which comes from Goldstone boson loop contributions to the
relevant Green's functions, is a next-to-leading order effect in the
$1/N_C$ expansion, and lies thus beyond the approximation
considered. We shall adopt the usual point of view that the estimates
furnished by this type of approach corresponds to the values of the
low-energy constants at the typical scale, say $\Lambda_H$, set by
these resonance states~\cite{Ecker:1989te}.

Approaches which do not rely on a resonance Lagrangian have been used
before at the ${\cal O}(p^6)$ level for two-point functions
\cite{Donoghue:1992ih,Knecht:1994ug,Golowich:1997kn}.  
In these studies, the relevant low-energy constants have often been
expressed through (superconvergent) dispersion integrals of the
corresponding spectral densities, which were then evaluated using
available data. It seems difficult to follow similar lines in the case
of three-point or higher Green's functions. Not only are their
analyticity properties far more complicated, but the corresponding
spectral densities are in general not experimental observables.
Studies of three-point functions similar to the lines we follow here
can be found in
Refs. \cite{Moussallam:1994at,Moussallam95,Moussallam97}, although the
discussion of their short-distance properties is less complete than
the one presented below.

In this article, we shall concentrate on a certain subset of ${\cal
O}(p^6)$ low-energy constants contributing to ${\cal L}_{(4,1)}$ and
${\cal L}_{(6,0)}$, and corresponding to the three-point functions
$\VAP$, $\VVP$ and $\AAP$ (see the beginning of section 2 for the
precise definitions). There are several reasons for this specific
choice. First, these correlators have been considered before in the
literature, so that quite some information concerning them is already
available. Second, although the approach that we shall follow here is,
in principle, applicable to other correlation functions as well, these
Green's functions have a certain degree of simplicity, which makes
them particularly valuable for illustrating our point of
view. Finally, these Greens functions also play a role in the
evaluation of some of the counterterms that arise in the calculation
of electromagnetic contributions to the pseudoscalar masses
\cite{Urech} or of radiative corrections to semileptonic decays of the
pseudoscalar mesons \cite{sirlin78,KNinprep} within an effective
Lagrangian framework~\cite{Knecht:2000ag}.

The remaining material of the present article is organized as follows.
In section 2, we define the relevant QCD Green's function and study
their long-distance properties in the chiral limit and at leading
order in the $1/N_C$ expansion. In particular, we identify the
low-energy constants related to these correlators. Section 3 is
devoted to an extensive discussion of the leading short-distance
properties of these Green's functions within the same framework. In
section 4, we construct some simple ans\"atze, in terms of a finite
number of narrow resonances, which correctly reproduce the
short-distance constraints. These are used to determine the
corresponding low-energy constants in section 5. We then compare our
results with those obtained from a Lagrangian involving
resonances~\cite{Prades} which has often been employed in the
literature to estimate the low-energy constants at order $p^6$
(section 6), and point out that this resonance Lagrangian does not
correctly incorporate the necessary short-distance properties (section
7). In section 8, we present several applications. Conclusions and
additional discussions can be found in section 9. Appendix A contains
some technical details relevant for the discussion in section 4 and
Appendix B gives the expression of the resonance Lagrangian of
Ref.~\cite{Prades}.


\section{Long-distance properties from chiral symmetry}
\renewcommand{\theequation}{\arabic{section}.\arabic{equation}}
\setcounter{equation}{0}

We consider, in the three flavour chiral limit, the momentum space QCD
three-point functions
\bea
(\Pi_{V\!AP})_{\mu\nu}^{abc}(p,q) = \int d^4x \int d^4y e^{i(p \cdot x + 
q \cdot y)} \langle 0 \vert T \{ V_\mu^a(x) A_\nu^b(y) P^c(0) \} \vert 0
\rangle \, , \lbl{VAPdef}
\nonumber
\eea
\bea
(\Pi_{VVP})_{\mu\nu}^{abc}(p,q) = \int d^4x \int d^4y e^{i(p \cdot x + 
q \cdot y)} \langle 0 \vert T \{ V_\mu^a(x) V_\nu^b(y) P^c(0) \} \vert 0
\rangle \, , \lbl{VVPdef}
\nonumber
\eea
\bea
(\Pi_{AAP})_{\mu\nu}^{abc}(p,q) = \int d^4x \int d^4y e^{i(p \cdot x + 
q \cdot y)} \langle 0 \vert T \{ A_\mu^a(x) A_\nu^b(y) P^c(0) \} \vert 0
\rangle \, , \lbl{AAPdef}
\eea
involving the octet vector and axial currents, 
\bea
V_\mu^a(x) & = & (\bar \psi \gamma_\mu {\lambda^a \over 2} \psi)(x) \, 
, \nonumber \\
A_\mu^a(x) & = & (\bar \psi \gamma_\mu \gamma_5 {\lambda^a \over 2}
\psi)(x)\, ,
\nonumber
\eea
as well as the octet pseudoscalar density
\bea
P^a(x)  =  (\bar \psi i \gamma_5 {\lambda^a \over 2} \psi)(x) \, .
\nonumber
\eea
These three-point functions satisfy the following chiral Ward identities,
\bea
p^{\mu}(\Pi_{V\!AP})_{\mu\nu}^{abc}(p,q) &=& 
\langle{\overline\psi}\psi\rangle_0 f^{abc}\,\Bigg[\frac{q_{\nu}}{q^2}\,-\,
\frac{(p+q)_{\nu}}{(p+q)^2}\Bigg]\, ,
\nonumber\\
&&
\nonumber\\
q^{\nu}(\Pi_{V\!AP})_{\mu\nu}^{abc}(p,q) &=&
\langle{\overline\psi}\psi\rangle_0 f^{abc}\,\frac{(p+q)_{\mu}}{(p+q)^2}\, ,
\nonumber
\eea
\bea
p^{\mu}(\Pi_{VVP})_{\mu\nu}^{abc}(p,q)=0\, ,  &&
q^{\nu}(\Pi_{VVP})_{\mu\nu}^{abc}(p,q)=0\, ,
\nonumber\\
&&
\nonumber\\
p^{\mu}(\Pi_{AAP})_{\mu\nu}^{abc}(p,q)=0\, ,  && 
q^{\nu}(\Pi_{AAP})_{\mu\nu}^{abc}(p,q)=0\, , 
\lbl{WardXXP}
\eea
where $\langle{\overline\psi}\psi\rangle_0$ denotes the single flavour
bilinear quark condensate in the chiral limit.  The general solution
of these Ward identities, taking into account the invariances of QCD
under $SU(3)_V$, parity and time reversal transformations (the latter
being responsible for the absence of structures of the type $d^{abc}$
in the case of $\Pi_{V\!AP}$, or of structures of the type $f^{abc}$
in the cases of $\Pi_{VVP}$ and $\Pi_{AAP}$), read
\bea
(\Pi_{V\!AP})_{\mu\nu}^{abc}(p,q) &=&
f^{abc}\,\Bigg\{
\langle{\overline\psi}\psi\rangle_0 \left[
\frac{(p+2q)_\mu q_\nu}{q^2 (p+q)^2} - \frac{\eta_{\mu\nu}}{(p+q)^2}\right] 
\nonumber \\
&&
+
P_{\mu\nu}(p,q) \F(p^2,q^2,(p+q)^2) 
+ Q_{\mu\nu}(p,q) \G(p^2,q^2,(p+q)^2) \Bigg\}\, , 
\nonumber\\
(\Pi_{VVP})_{\mu\nu}^{abc}(p,q) & = &
\eps_{\mu\nu\alpha\beta}p^{\alpha}q^{\beta} 
\,d^{abc}\, \H_V(p^2,q^2,(p+q)^2)
 \, , 
\nonumber \\ 
&&\nonumber\\
(\Pi_{AAP})_{\mu\nu}^{abc}(p,q) & = &
\eps_{\mu\nu\alpha\beta}p^{\alpha}q^{\beta} 
\,d^{abc}\, \H_A(p^2,q^2,(p+q)^2) 
 \, .\lbl{XXPsol} 
\eea 
Here, $\eta_{\mu\nu}$ denotes the metric tensor in flat Minkowski
space with signature $(+,-,-,-)$ and we use the conventions
$\eps_{0123} = 1$ for the totally antisymmetric tensor
$\eps_{\mu\nu\rho\sigma}$ and $\gamma_5 = i \gamma^0 \gamma^1
\gamma^2 \gamma^3$.  The transverse tensors $P_{\mu\nu}$ and 
$Q_{\mu\nu}$ are defined by
\bea
P_{\mu\nu}(p,q) = q_\mu p_\nu - (p \cdot q) \eta_{\mu\nu} \, , \ \ 
Q_{\mu\nu}(p,q) = p^2 q_\mu q_\nu + q^2 p_\mu p_\nu - (p \cdot q)
p_\mu q_\nu - p^2 q^2 \eta_{\mu\nu} \, .  
\nonumber
\eea
Due to Bose-Einstein symmetry, the invariant functions $\H_{V,A}$ have
the property  
\bea \label{Bose_symmetry} 
\H_{V,A}(p^2,q^2,(p+q)^2) = \H_{V,A}(q^2,p^2,(p+q)^2)
\, .  
\eea

The behaviour of these invariant functions at small momentum transfers
is constrained by the presence of singularities arising from 
Goldstone boson intermediate states. Here we are
interested in the limit where the number of colours $N_C$ becomes
infinite. In this limit, the contributions from one-particle
intermediate states dominate, so that at low energies we only need to
keep the corresponding Goldstone boson poles and the polynomial terms
involving the counterterms. In the even intrinsic parity case, we use
the basis of Ref. \cite{ChPT_p6} for the ${\cal O}(p^6)$
counterterms, and we obtain
\bea
\F^{\XPT}(p^2,q^2,(p+q)^2) & = & \frac{4
\langle{\overline\psi}\psi\rangle_0}{F_0^2 (p+q)^2} \Big[  L_9 +  
 L_{10} \nonumber \\
&&\qquad + (C_{78} - \frac{5}{2}C_{88} - C_{89} + 3C_{90}) p^2 
+ (C_{78} - 2C_{87} + \frac{1}{2}C_{88}) q^2 
\nonumber\\
&&\qquad\qquad
+ (C_{78} + 4C_{82} - \frac{1}{2}C_{88})(p+q)^2
\Big] \,+\cdots\, , \nonumber \\ 
\G^{\XPT}(p^2,q^2,(p+q)^2) & = & \frac{4
\langle{\overline\psi}\psi\rangle_0}{F_0^2 q^2 (p+q)^2} \Big[  L_9 
\nonumber \\
&&+ 2( -C_{88} +  C_{90}) p^2 
+ (2C_{78} - C_{89} + C_{90}) q^2 - 2  C_{90} (p+q)^2
\Big] \,+\cdots\, ,\nonumber \\
&& \lbl{FG_ChPT}  
\eea
where the ellipses stand for higher order contributions.
For the two correlators of odd intrinsic parity, we use the counterterm 
Lagrangian of Ref. \cite{Fearing_Scherer}, in terms of which we find
\bea
\H_V^{ChPT}(p^2,q^2,(p+q)^2) & = & 
-\,\frac{\langle{\overline\psi}\psi\rangle_0}{F_0^2 (p+q)^2} \Bigg[ 
-\, \frac{N_C}{8 \pi^2} \nonumber + 
(4 A_2 - 16 A_3)( p^2 + q^2) \nonumber \\
&&\qquad\quad + (- 4 A_2 + 8 A_3 + 16 A_4) (p+q)^2 \Bigg] \,+\cdots\,
, \nonumber \\ 
\H_A^{ChPT}(p^2,q^2,(p+q)^2) & = & 
-\,\frac{\langle{\overline\psi}\psi\rangle_0}{F_0^2 (p+q)^2} \Bigg[ 
-\, \frac{N_C}{24 \pi^2} \nonumber + 
(4 A_{11} + 4 A_{23} - 16 A_{24})( p^2 + q^2) \nonumber \\
&&\qquad\quad 
+ (- 12 A_{11} - 16 A_{17} - 4 A_{23} + 8 A_{24} + 16 A_{25}) (p+q)^2 \Bigg] 
+\cdots\, . \nonumber\\
&&\lbl{H_VA_ChPT} 
\eea
We have thus identified the set of low-energy constants that describe the 
long-distance behaviour of the  $\VAP$, $\VVP$ and $\AAP$ three-point
functions.  


\section{Short-distance analysis}
\label{sec:short_distance}
\renewcommand{\theequation}{\arabic{section}.\arabic{equation}}
\setcounter{equation}{0}

We next study the properties of the $\VAP$, $\VVP$ and $\AAP$
correlators at short distances.  These will be conditioned by the fact
that the three Green's functions under consideration are order
parameters of chiral symmetry. Therefore, they vanish to all orders in
perturbative QCD in the chiral limit, so that their behaviour at short
distances is smoother than expected from naive power counting
arguments.  Two limits are of interest. In the first case, the two
momenta become simultaneously large, which in position space describes
the situation where the space-time arguments of the three operators
tend towards the same point at the same rate. Our analysis will be
restricted to the leading terms~\footnote{In the case of the $\VAP$
correlator, the subleading term in the short-distance expansion,
involving the mixed condensate, can be found in
Ref. \cite{Moussallam97}.}, and the expressions below hold up to
corrections of order $\order(\alpha_s)$. We obtain
\bea
\lim_{\lambda\to\infty}(\Pi_{V\!AP})_{\mu\nu}^{abc}(\lambda p,\lambda q)
&=& \frac{\langle{\overline\psi}\psi\rangle_0}{\lambda^2}\,f^{abc}\,
\frac{1}{p^2q^2(p+q)^2}\,\bigg\{\,
p^2(p+2q)_{\mu}q_{\nu}-\eta_{\mu\nu}p^2q^2
\nonumber\\
&&
\qquad
+\frac{1}{2}[p^2-q^2-(p+q)^2]P_{\mu\nu}-Q_{\mu\nu}\,
\bigg\}\,+\,{\cal O}\left(\frac{1}{\lambda^4}\right)\, ,
\nonumber \\
\lim_{\lambda\to\infty}(\Pi_{VVP})_{\mu\nu}^{abc}(\lambda p,\lambda q)
&=& -\,\frac{\langle{\overline\psi}\psi\rangle_0}{2\lambda^2}\,d^{abc}\,
\epsilon_{\mu\nu\alpha\beta}p^{\alpha}q^{\beta}\,
\frac{q^2+p^2+(p+q)^2}{p^2q^2(p+q)^2}
\,+\,{\cal O}\left(\frac{1}{\lambda^4}\right) \, , 
\nonumber\\
\lim_{\lambda\to\infty}(\Pi_{AAP})_{\mu\nu}^{abc}(\lambda p,\lambda q)
&=& -\,\frac{\langle{\overline\psi}\psi\rangle_0}{2\lambda^2}\,d^{abc}\,
\epsilon_{\mu\nu\alpha\beta}p^{\alpha}q^{\beta}\,
\frac{q^2+p^2-(p+q)^2}{p^2q^2(p+q)^2}
\,+\,{\cal O}\left(\frac{1}{\lambda^4}\right) \, . 
\eea
One concludes that
\bea
\lim_{\lambda \to \infty} \F((\lambda p)^2, (\lambda q)^2, (\lambda p
+\lambda q)^2) &=& {1\over 2
\lambda^4}\,\langle{\overline\psi}\psi\rangle_0\, 
{p^2 - q^2 - (p+q)^2 \over
p^2 q^2 (p+q)^2} + \order\left({1\over \lambda^6}\right) \, ,
\lbl{F_OPE} \\ 
%
\lim_{\lambda \to \infty} \G((\lambda p)^2, (\lambda q)^2, (\lambda p
+ \lambda q)^2) &=& - {1\over \lambda^6} 
\,\frac{\langle{\overline\psi}\psi\rangle_0}{p^2 q^2 (p+q)^2} + 
 \order\left({1\over \lambda^8}\right) \, , \lbl{G_OPE} 
\eea
and
\be
\lim_{\lambda \to \infty} \H_{V,A}((\lambda p)^2, (\lambda q)^2,
(\lambda p + \lambda q)^2) \ =\ -\,\frac{1}{2 \lambda^4}
\,\langle{\overline\psi}\psi\rangle_0\,
\frac{p^2 + q^2 \pm (p+q)^2}{p^2 q^2 (p+q)^2} 
+ \order\left({1\over \lambda^6}\right) \, . \lbl{H_OPE}
\ee
Notice that since the $\langle{\overline\psi}\psi\rangle_0$ condensate
and the pseudoscalar density $P^a(x)$ have the same anomalous
dimensions, the leading short-distance behaviour exhibited in these
expressions is canonical, the corresponding Wilson coefficients have
no anomalous dimensions.

The second situation of interest corresponds to the case where the
relative distance between only two of the three operators involved
becomes small. It so happens that the corresponding behaviours in
momentum space involve, apart from the correlator $\langle A P\rangle$ 
which, in the chiral limit, is saturated by the single-pion 
intermediate state,
\bea
\int d^4 x e^{ip \cdot x}
\langle 0 \vert T \{ A_\mu^a(x) P^b(0) \} \vert 0 \rangle =
\delta^{ab} \langle{\overline\psi}\psi\rangle_0 \, {p_\mu \over p^2} \, , 
\nonumber 
\eea   
the two-point function $\langle VT\rangle$ of the vector current and
the antisymmetric tensor density, 
\bea
\delta^{ab}(\Pi_{VT})_{\mu\rho\sigma}(p)\,=\, 
\int d^4x e^{ip \cdot x}
\langle 0 \vert T \{ V_\mu^a(x) 
({\overline\psi}\,\sigma_{\rho\sigma}\frac{\lambda^b}{2}\,\psi)(0)\}\vert
0\rangle \, ,  
\nonumber
\eea
with $\sigma_{\rho\sigma}={i\over 2}[\gamma_{\rho},\gamma_{\sigma}]$
(the similar correlator between the axial current and the tensor
density vanishes as a consequence of invariance under charge conjugation). 
Conservation of the vector current and invariance under parity then give
\bea
(\Pi_{VT})_{\mu\rho\sigma}(p)\,=
\,(p_{\rho}\eta_{\mu\sigma}-p_{\sigma}\eta_{\mu\rho})\,\Pi_{VT}(p^2)
\,. 
\nonumber
\eea
The leading short-distance behaviour of this two-point
function reads (see also \cite{Craigie:1982jx})
\bea \label{VT_OPE} 
\lim_{\lambda \to \infty}\Pi_{VT}((\lambda p)^2)\,=\,
-\,\frac{1}{\lambda^2}
\,\frac{\langle{\overline\psi}\psi\rangle_0}{p^2}
\,+\,{\cal O}\left(\frac{1}{\lambda^4}\right)\,. 
\eea
For the $\VAP$ correlator, we then find 
\bea
\lim_{\lambda \to \infty}
(\Pi_{V\!AP})_{\mu\nu}^{abc}(\lambda p,q-\lambda p)\ =
\ -\,\frac{1}{\lambda}\,f^{abc}\,\langle{\overline\psi}\psi\rangle_0
\,\frac{p_{\mu}q_{\nu}+p_{\nu}q_{\mu}-(p\cdot q)\eta_{\mu\nu}}{p^2q^2}
\,+\,{\cal O}\left(\frac{1}{\lambda^2}\right)\, ,
\lbl{VAPope1}
\eea
\bea
\lim_{\lambda \to \infty}
(\Pi_{V\!AP})_{\mu\nu}^{abc}(\lambda p,q)\ =
\ \frac{1}{\lambda}\,f^{abc}\,\langle{\overline\psi}\psi\rangle_0\,
\frac{p_{\mu}q_{\nu}}{p^2q^2}
\,+\,{\cal O}\left(\frac{1}{\lambda^2}\right)\, ,
\eea
and
\bea
\lim_{\lambda \to \infty}
(\Pi_{V\!AP})_{\mu\nu}^{abc}(p,\lambda q)\ =
\ \,\frac{1}{\lambda}\,f^{abc}\,\frac{p_{\nu}q_{\mu}-(p\cdot
q)\eta_{\mu\nu}}{q^2} 
\,\Pi_{VT}(p^2)
\,+\,{\cal O}\left(\frac{1}{\lambda^2}\right)\, .
\eea
For the $\langle VVP\rangle$ correlator we obtain the results 
\bea
\lim_{\lambda \to \infty}
(\Pi_{VVP})_{\mu\nu}^{abc}(\lambda p,q-\lambda p)\ =
\ -\,\frac{1}{\lambda}\,d^{abc}\,\langle{\overline\psi}\psi\rangle_0\,
\epsilon_{\mu\nu\rho\sigma}p^{\rho}q^{\sigma}\,\frac{1}{p^2q^2}
\,+\,{\cal O}\left(\frac{1}{\lambda^2}\right) \, , 
\eea
\bea
\lim_{\lambda \to \infty}
(\Pi_{VVP})_{\mu\nu}^{abc}(\lambda p,q)\ =
\ \frac{1}{\lambda}\,d^{abc}\,\epsilon_{\mu\nu\rho\sigma}
\,\frac{p^{\rho}q^{\sigma}}{p^2}\,\Pi_{VT}(q^2)
\,+\,{\cal O}\left(\frac{1}{\lambda^2}\right)\, ,
\eea
and for the  $\AAP$ correlator,
\bea
\lim_{\lambda \to \infty}
(\Pi_{AAP})_{\mu\nu}^{abc}(\lambda p,q-\lambda p)\ =
\ -\,\frac{1}{\lambda}\,d^{abc}\,\langle{\overline\psi}\psi\rangle_0\,
\epsilon_{\mu\nu\rho\sigma}p^{\rho}q^{\sigma}\,\frac{1}{p^2q^2}
\,+\,{\cal O}\left(\frac{1}{\lambda^2}\right)\, ,
\eea
\bea
\lim_{\lambda \to \infty}
(\Pi_{AAP})_{\mu\nu}^{abc}(\lambda p,q)\ =
\ {\cal O}\left(\frac{1}{\lambda^2}\right)\, .
\eea
In terms of the invariant functions $\F$ and $\G$, the constraint
\rf{VAPope1} yields 
\bea
\lim_{\lambda \to \infty}
\F((\lambda p)^2,(q-\lambda p)^2,q^2) &=&
\frac{\langle{\overline\psi}\psi\rangle_0}{\lambda^2 p^2}\, 
\left[\F^{(0)}(q^2)\,+\, \frac{1}{\lambda}\,\frac{p\cdot q}{p^2}
\,\F^{(1)}(q^2)
\,+\,\order\left({1\over \lambda^2}\right)\right] \, , 
\nonumber\\
\lim_{\lambda \to \infty}
\G((\lambda p)^2,(q-\lambda p)^2,q^2) &=&
\frac{\langle{\overline\psi}\psi\rangle_0}{(\lambda^2 p^2)^2}\, 
\left[\G^{(0)}(q^2)\,+\, \frac{1}{\lambda}\,\frac{p\cdot q}{p^2}
\,\G^{(1)}(q^2)
\,+\,\order\left({1\over \lambda^2}\right)\right] \, , 
\lbl{VAPope2}
\eea
together with
\bea
\F^{(0)}(q^2)-\G^{(0)}(q^2) &=& \frac{1}{q^2} \, , 
\nonumber\\
\F^{(1)}(q^2)-\G^{(1)}(q^2)+\G^{(0)}(q^2)&=& \frac{2}{q^2}\,.
\lbl{VAPope3}
\eea
Finally, the following properties
\bea
\lim_{\lambda \to \infty}\F((\lambda p)^2, q^2, (q+\lambda p)^2)
&=& \order\left({1\over \lambda^3}\right) \, , 
\nonumber\\
\lim_{\lambda \to \infty}\G((\lambda p)^2, q^2, (q+\lambda p)^2)
&=& \order\left({1\over \lambda^4}\right) \, , 
\lbl{VAPope4}
\eea
and
\bea
\lim_{\lambda \to \infty}\F(p^2,(\lambda q)^2, (p+\lambda q)^2)
&=&\frac{1}{\lambda^2}\,\frac{1}{q^2}\,\Pi_{VT}(p^2) \,+\,
\order\left({1\over \lambda^3}\right) \, , 
\nonumber\\
\lim_{\lambda \to \infty}\G(p^2,(\lambda q)^2, (p+\lambda q)^2)
&=& \order\left({1\over \lambda^4}\right)\,, 
\lbl{VAPope5}
\eea
must also be satisfied. For the invariant functions $\H_V$ and $\H_A$
we obtain the constraints 
\be
\lim_{\lambda \to \infty} \H_{V,A}((\lambda p)^2, (q-\lambda p)^2,q^2) 
= -\,\frac{1}{\lambda^2}
\,\langle{\overline\psi}\psi\rangle_0\,
\frac{1}{p^2 q^2} 
+ \order\left({1\over \lambda^3}\right) \, , \label{H_V/A_OPE}
\ee
\bea
\lim_{\lambda \to \infty} \H_{V}((\lambda p)^2,q^2,(q+\lambda p)^2)
&=& \frac{1}{\lambda^2}\,\frac{1}{p^2}\,\Pi_{VT}(q^2)
+ \order\left({1\over \lambda^3}\right) \, , \label{H_V_x_to_0} \\
\lim_{\lambda \to \infty} \H_{A}((\lambda p)^2,q^2,(q+\lambda p)^2)
&=& \order\left({1\over \lambda^3}\right) \, . \label{H_A_x_to_0}
\eea


\section{The intermediate energy region}
\renewcommand{\theequation}{\arabic{section}.\arabic{equation}} 
\setcounter{equation}{0}

In this section, we shall construct representations of the invariant
functions which describe the correlators $\VAP$, $\VVP$, $\AAP$ and
$\langle VT \rangle$ in the intermediate energy region, dominated by
the resonances, and which reproduce the short-distance constraints
studied in the preceding section. Finding the general structure of the
invariant functions $\F$, $\G$ and $\H_{V,A}$ is of course far beyond
our present possibilities. As discussed in the introduction, we shall
therefore work in the framework of the large-$N_C$ approximation to
QCD and assume, in addition, that already a finite number of
resonances will give a satisfactory description of these
correlators.  

We first consider the case where only a single resonance is retained
in each channel, assuming that in the pseudoscalar one only the massless 
Goldstone bosons
need to be kept. The corresponding lowest meson dominance (LMD)
ans\"atze for the invariant functions are constructed such that the
resulting expressions agree with the short-distance constraints
\rf{F_OPE} -- \rf{H_OPE}. This is rather straightforward, and in the
case of the $\VAP$ correlator, the result reads
\cite{Moussallam97} 
\bea
\F^{LMD}(p^2,q^2,(p+q)^2) & = & 
\frac{\langle{\overline\psi}\psi\rangle_0}{2} {p^2 - q^2 - (p+q)^2 + 2
a\over (p^2 - M_V^2) (q^2 - M_A^2) (p+q)^2} \, , 
\nonumber \\ 
\G^{LMD}(p^2,q^2,(p+q)^2) & = & 
\langle{\overline\psi}\psi\rangle_0\,
\frac{- q^2 + b }{(p^2 - M_V^2) (q^2 -M_A^2) q^2 (p+q)^2} \, . 
\label{FG_LMD} 
\eea
The constants $a$ and $b$ in Eq.~(\ref{FG_LMD}) can be determined
as follows. As shown in Ref.~\cite{Moussallam97}, one can relate the
$\langle VA | \pi\rangle$ vertex function $\Gamma_{VA}$ (for the definitions 
and a discussion of some properties of the vertex functions associated to the 
three point functions under study, we refer the reader to Appendix A)  to
the two-point correlator $\langle VV - AA\rangle$ via a 
low-energy theorem. From the
two Weinberg sum rules~\cite{Weinberg_SR} one obtains in this way the
relation 
\be \label{a_b} 
a - b = - (M_V^2 + M_A^2) \, . 
\ee
The constant $b$ can be fixed by requiring that the vector form factor 
of the pion $F_V^\pi(q^2)$, defined by 
\be
\langle \pi^a(p^\prime) | V_\mu^b(0) | \pi^c(p) \rangle 
=  i f^{abc} (p^\prime + p)_\mu F_V^\pi(q^2) \, , \quad q = p -
p^\prime \, ,  \label{def_F_V}
\ee
satisfies an unsubtracted dispersion relation~\cite{Ecker:1989yg}
(there are theoretical arguments~\cite{pion_formfactor,Brodsky_Lepage} 
in favour of a $1/q^2$ fall-off of this form factor for large momentum
transfer). Since
\bea
F_V^\pi(q^2) \equiv 1 + {q^2 \over 2 \qbarq}\, \lim_{(q-p)^2 \to 0}\,
 \lim_{p^2 \to 0} (q-p)^2 p^2 \G(q^2,(q-p)^2,p^2) \, , 
\nonumber 
\eea
we obtain with the function $\G^{LMD}$ from Eq.~(\ref{FG_LMD}) the
result  
\bea
F_V^{\pi,LMD}(q^2) = 1 - \frac{b}{2M_A^2}\,\frac{q^2}{q^2-M_V^2}\, ,
\eea
and thus
\be \label{b} 
b = 2 M_A^2 \, .  
\ee
Combining Eqs.~(\ref{a_b}) and (\ref{b}) then gives
\be
a = M_A^2 - M_V^2 \, . 
\ee

We note that the argument given in Ref.~\cite{Moussallam97} that the
same result for the constant $b$ can also be obtained by enforcing the
correct short-distance behaviour of the $\langle V P \vert \pi
\rangle$ vertex function $\Gamma_{VP}$ is not correct. 
In Appendix~\ref{app:OPE} we
sketch the derivation of the operator product expansion of this vertex
function, with the result given in Eq.~(\ref{OPE_JJpi}). The LMD
ansatz will reproduce the subleading term in the OPE provided $b = 4
M_A^2 / 3$, which implies with Eq.~(\ref{a_b}) $a = - M_V^2 + M_A^2 /
3$. Thus, with the LMD ansatz for the invariant function $\G$ from
Eq.~(\ref{FG_LMD}) it is not possible to reproduce at the same time 
the requirements from the asymptotic behaviour of the form factor 
$F_V^\pi(q^2)$ and from the subleading terms in the OPE for the 
$\langle V P | \pi\rangle$ vertex function. 
The subleading term of the OPE of $\langle V A | \pi\rangle$ is correctly 
reproduced if Eq.~(\ref{a_b}) holds. In what follows, we shall 
understand that the LMD approximation for the $\VAP$ correlator corresponds 
to the ansatz (\ref{FG_LMD}) together with the choice $b=2M_A^2$.

For the $\langle VT\rangle$ two-point functions, only $J^{PC}=1^{--}$
vector mesons contribute.  In the LMD approximation, the leading
short-distance behaviour then fixes everything but the mass of the
lowest vector resonance,
\bea
\Pi_{VT}^{LMD}(p^2) \,=\, -\,\langle{\overline\psi}\psi\rangle_0\,
\frac{1}{p^2-M_V^2} \, .  
\eea

It is quite remarkable that with this simple ansatz all the remaining
leading short-distance constraints explicited in the previous section
are met. In particular, for the quantities introduced in
Eq. \rf{VAPope2} we obtain
\bea
\F^{(0),LMD}(q^2)\,=\,0 \, , && \F^{(1),LMD}(q^2)\,=\,\frac{1}{q^2} \, 
, 
\nonumber\\
\G^{(0),LMD}(q^2)\,=\,-\,\frac{1}{q^2} \, , &&
\G^{(1),LMD}(q^2)\,=\,-\,\frac{2}{q^2} \, , 
\nonumber
\eea
which satisfy \rf{VAPope3}. Note, however, that the LMD ans\"atze for
$\F$ and $\G$ in Eq.~(\ref{FG_LMD}) do not correctly reproduce the
subleading terms in the OPE for the correlator $\VAP$ given in
Ref.~\cite{Moussallam97}.

The LMD ansatz for the invariant functions ${\cal H}_{V,A}$
reads~\footnote{The LMD ansatz for the $\VVP$ three-point function was
given in Refs. \cite{Moussallam95,Pi_ll}.} 
\bea
\H_V^{LMD}(p^2,q^2,(p+q)^2) & = &
-\,\frac{\langle{\overline\psi}\psi\rangle_0}{2}\, {p^2 + q^2 + 
(p+q)^2 - c_V \over (p^2 - M_V^2) (q^2 - M_V^2) (p+q)^2} , \quad c_V =
{N_C \over 4\pi^2} {M_V^4 \over F_0^2} ,  
\nonumber\\
\H_A^{LMD}(p^2,q^2,(p+q)^2) & = &
-\,\frac{\langle{\overline\psi}\psi\rangle_0}{2}\, {p^2 + q^2 - 
(p+q)^2 - c_A \over (p^2 - M_A^2) (q^2 - M_A^2) (p+q)^2} , \quad c_A =
{N_C \over 12\pi^2} {M_A^4 \over F_0^2} , 
\eea
where $c_{V,A}$ are fixed by the Wess-Zumino-Witten anomaly~\cite{WZW}
term.  Again, these simple ans\"atze fulfill all the remaining leading
short-distance requirements worked out in the preceding section. We
note that for the vertex functions $\Gamma_{VV}$ and $\Gamma_{AA}$
also the subleading terms in the OPE in Eq.~(\ref{OPE_JJpi}) are
reproduced by these ans\"atze.    

As already pointed out, it is sometimes necessary to generalize the
ans\"atze for the invariant functions given above by including more
than one resonance in each channel. This might be due to some
additional constraints that are imposed on the Green's functions or in
order to better reproduce experimental data involving resonances, as
was argued in Ref.~\cite{Moussallam95}. 
If we include, for instance, one additional vector resonance,
the expressions for the invariant functions $\F, \G, \Pi_{VT}$ and
$\H_V$ read as follows ($\H_A$ remains, of course, unchanged)
\bea
\F^{LMD+V}(p^2,q^2,(p+q)^2) & = & 
\frac{\langle{\overline\psi}\psi\rangle_0}{2} {p^2 \left[ p^2 - q^2 -
(p+q)^2 \right] + P_F(p^2,q^2,(p+q)^2) \over (p^2 - M_{V_1}^2) (p^2 -
M_{V_2}^2) (q^2 - M_A^2) (p+q)^2} \, , \nonumber \\ 
\G^{LMD+V}(p^2,q^2,(p+q)^2) & = & 
\langle{\overline\psi}\psi\rangle_0\,
{- p^2 q^2 + P_G(p^2, q^2, (p+q)^2) \over (p^2 - M_{V_1}^2) (p^2 -
M_{V_2}^2) (q^2 - M_A^2) q^2 (p+q)^2}
\, , \nonumber \\ 
\Pi_{VT}^{LMD+V}(p^2) & = &  -\,\langle{\overline\psi}\psi\rangle_0\, 
{ p^2 + c_{VT} \over (p^2-M_{V_1}^2) (p^2-M_{V_2}^2) } \, ,
\nonumber \\ 
\H_V^{LMD+V}(p^2,q^2,(p+q)^2) & = &
-\,\frac{\langle{\overline\psi}\psi\rangle_0}{2}\, { p^2 q^2 \left[
p^2 + q^2 + (p+q)^2 \right] + P_H^V(p^2,q^2,(p+q)^2) \over (p^2 -
M_{V_1}^2) (p^2 - M_{V_2}^2) (q^2 - M_{V_1}^2) (q^2 - M_{V_2}^2)
(p+q)^2} \, , \nonumber \\
& & \label{FGHVA_LMD+V} 
\eea
where
\bea
P_F(p^2, q^2, (p+q)^2) & = & f_1 p^2 + f_2 q^2 + f_3 (p+q)^2 + f_4 \,
, \label{polynom_F} \nonumber \\
P_G(p^2, q^2, (p+q)^2) & = & g_1 p^2 + g_2 q^2 + g_3 (p+q)^2 + g_4 \,
, \label{polynom_G} \nonumber \\
P_H^V(p^2, q^2, (p+q)^2) & = & \hv_1 (p^2 + q^2)^2 + \hv_2 p^2 q^2 +
\hv_3 (p^2 + q^2) (p+q)^2 + \hv_4 (p+q)^4 \nonumber \\
& &+ \hv_5 (p^2 + q^2) + \hv_6
(p+q)^2 + \hv_7 \, . \nonumber  \label{polynom_HV} 
\eea
The coefficients that appear in the polynomials $P_F, P_G$ and $P_H^V$
have to fulfill the following relations 
\bea
f_2 + f_3 & = & -2 c_{VT} \, , \nonumber \\
\hv_1 + \hv_3 + \hv_4 & = & 2 c_{VT} \, , 
\eea
in order to reproduce all short-distance constraints from the operator
product expansion given in the previous section.  Furthermore, the
Wess-Zumino-Witten anomaly determines
\bea
\hv_7 = - {N_C \over 4 \pi^2} {M_{V_1}^4 M_{V_2}^4 \over F_0^2} \, . 
\nonumber 
\eea
Using  the low-energy theorem that relates $\Gamma_{VA}$ to 
$\langle VV - AA \rangle$ one obtains, from the LMD+V
ansatz for the latter correlator~\cite{Knecht_deRafael},  the relations
\bea
f_1 + f_2 & = & 2 \left( g_1 + g_2 - M_{V_1}^2 - M_{V_2}^2 - M_A^2
\right) \, ,  \nonumber \\
f_4 & = & 2 \left( g_4 + M_{V_1}^2 M_{V_2}^2 + M_A^2 (M_{V_1}^2 +
M_{V_2}^2) - { 4\pi \alpha_s \langle {\overline\psi}\psi \rangle_0^2
\over F_0^2 } \right) \,  . 
\eea
The first of these conditions also guarantees that the next-to-leading 
short-distance behaviour of $\Gamma_{VA}$ is correctly reproduced.
The vector form factor of the pion now reads
\bea
F_V^{\pi,LMD+V}(q^2) = 1 - \frac{q^2}{2M_A^2}\,\frac{g_1q^2+g_4}
{(q^2-M_{V_1}^2)(q^2-M_{V_2}^2)}\, .
\eea
Requiring that it behaves 
like $1/q^2$ for large $q^2$, leads to the relation
\be \label{g1_Formfactor} 
g_1 = 2 M_A^2 \, . 
\ee 
On the other hand, in order to reproduce the subleading terms in the
OPE for $\Gamma_{VP}$, Eq.~(\ref{OPE_JJpi}), with the
LMD+V ansatz for $\G$ yields a constraint which is independent from the 
previous one,
\be
g_1 + g_3 = {4 \over 3} M_A^2 \, .   
\ee
Combining this equation with Eq.~(\ref{g1_Formfactor}), we
obtain the result 
\bea
g_3 = - {2 \over 3} M_A^2 \, . \nonumber
\eea
Note that in contrast to the LMD case above, we can simultaneously 
fulfill the requirements from the asymptotic behaviour of the form
factor $F_V^\pi(q^2)$ and from the subleading terms in the OPE for the
$\Gamma_{VP}$ vertex functions. 

Finally we note that the subleading terms in the OPE for the $\langle
VV | \pi \rangle$ vertex function $\Gamma_{VV}$ in
Eq.~(\ref{OPE_JJpi}) are reproduced by the LMD+V ansatz for $\H_V$
without leading to further constraints on the coefficients $h_i$.

Let us briefly mention the ansatz for $\H_V$ with one additional 
pseudoscalar resonance, discussed in Ref.~\cite{Moussallam95},  
\bea
\H_V^{LMD+P}(p^2,q^2,(p+q)^2) & = &
-\,\frac{\langle{\overline\psi}\psi\rangle_0}{2}\, { (p+q)^2 \left[
p^2 + q^2 + (p+q)^2 \right] + P_H^P(p^2,q^2,(p+q)^2) \over (p^2 -
M_V^2) (q^2 - M_V^2) ((p+q)^2 - M_P^2) (p+q)^2} \, , \nonumber \\
& & \label{H_V_LMD+P}  
\eea
where
\bea
P_H^P(p^2,q^2,(p+q)^2) & = & \hp_1 (p^2 + q^2) + \hp_2 (p+q)^2 + \hp_3
\, .  \nonumber  
\eea
The OPE leads to the condition
\be \label{OPE_hP1} 
\hp_1 = - M_P^2 \, , 
\ee
and the Wess-Zumino-Witten anomaly yields
\bea
\hp_3 & = & {N_C \over 4\pi^2} {M_V^4 M_P^2 \over F_0^2} \,
. \nonumber 
\eea

In contrast to the LMD case, the coefficients that appear in the
invariant functions are no longer fixed unambiguously by the leading
terms in the OPE. Here we have considered additional restrictions
arising from the next-to-leading short-distance behaviour of the
$\langle V A \vert \pi\rangle$, $\langle V P \vert \pi\rangle$ and
$\langle V V | \pi \rangle$ vertex functions.  Further information may
be gained from the study of vertex functions like $\langle V P \vert
a_1\rangle$, $\langle V V \vert \rho\rangle$, etc.  We shall not
pursue this interesting line of thought here. Other sources of
additional constraints might also be invoked, either from low-energy
physics, from subleading terms in the OPE of the three-point functions
\footnote{Actually, the addition of a single vector or pseudoscalar
resonance is still not sufficient in order to reproduce the next-to-leading
short-distance behaviour of $\VAP$ as given by Eq. (46) of Ref.
\cite{Moussallam97}.} or from processes involving resonances. We shall
illustrate this point below.


\section{Determination of low-energy constants}
\renewcommand{\theequation}{\arabic{section}.\arabic{equation}}
\setcounter{equation}{0}

In this section, we shall use the ans\"atze of the previous section in
order to obtain an evaluation of the low-energy constants involved in
the chiral expansion of the three correlators under study. This is
done upon performing the low-energy expansion of the invariant
functions $\F^{LMD}$, $\G^{LMD}$ and  $\H_{V,A}^{LMD}$ and by
subsequently matching it with the chiral expressions
\rf{FG_ChPT} and \rf{H_VA_ChPT}. 

We start with the sector of even intrinsic parity, i.e. with the
$\VAP$ correlator. For small momentum transfers, we can expand the
resonance propagators 
\be \label{expand_prop}
{1\over p^2 - M_R^2} = - {1\over M_R^2} \left[ 1 + \order\left( {p^2
\over M_R^2} \right) \right] \, ,
\ee
and obtain (in these expressions, we have taken $b=2M_A^2$, $a=M_A^2-M_V^2$, 
see the discussion after Eq. (\ref{FG_LMD})),
\bea
\lefteqn{ \F^{LMD}(p^2,q^2,(p+q)^2) =
\frac{\langle{\overline\psi}\psi\rangle_0}{(p+q)^2}\, \Bigg[ {1\over 
M_V^2} - {1\over M_A^2} } \nonumber \hspace{2cm} \\
&&+ {p^2 \over M_V^2 M_A^2} \left({M_A^2 \over M_V^2} - {1\over 
2}\right) + {q^2 \over M_V^2 M_A^2} \left({1\over 2} - {M_V^2 \over
M_A^2}\right) - {1\over 2} {(p+q)^2 \over M_V^2 M_A^2} + \order(p^4)
\Bigg] \, ,\nonumber  \label{F_LMD_low}  
\eea
\bea
\G^{LMD}(p^2,q^2,(p+q)^2) =
\frac{\langle{\overline\psi}\psi\rangle_0}{q^2 (p+q)^2}\, \Bigg[
{2\over 
M_V^2} + p^2 \left({2 \over M_V^4}\right) + q^2 \left({1 \over M_V^2
M_A^2}\right) + \order(p^4) \Bigg] \, ,\nonumber   \label{G_LMD_low} 
\eea
where $\order(p^4)$ includes all possible higher order polynomials in $p^2,
q^2, (p+q)^2$. Comparison with the expressions~\rf{FG_ChPT} from ChPT
yields the following solution (treating $SU(2)$ and 
$SU(3)$ together)
\bea
L_9^{LMD} & = & -{1\over 2} l_6^{LMD} = {1\over 2} {F_0^2 \over M_V^2}
\, ,\nonumber \\ 
L_{10}^{LMD} & = & l_5^{LMD} = - {1\over 4} {F_0^2 \over M_V^2} -
{1\over 4} {F_0^2 \over M_A^2} \, , 
\eea
and
\bea 
C_{78}^{LMD} & = & c_{44}^{LMD} = {3\over 8} {F_0^2 \over M_V^4} +
{3\over 8} {F_0^2 \over M_V^2 M_A^2} \, , \label{C_i_LMD_78}\nonumber \\
C_{82}^{LMD} & = & c_{47}^{LMD} = - {1\over 8} {F_0^2 \over M_V^4}
- {1\over 8} {F_0^2 \over M_V^2 M_A^2} \, , \nonumber\\
C_{87}^{LMD} & = & c_{50}^{LMD} = {1\over 8} {F_0^2 \over M_V^4} +
{1\over 8} {F_0^2 \over M_V^2 M_A^2} + {1\over 8} {F_0^2 \over M_A^4}
\, , \nonumber\\ 
C_{88}^{LMD} & = & c_{51}^{LMD} = - {1\over 4} {F_0^2 \over M_V^4}
\, , \nonumber\\  
C_{89}^{LMD} & = & c_{52}^{LMD} = {3\over 4} {F_0^2 \over M_V^4} +
{1\over 2} {F_0^2 \over M_V^2 M_A^2} \, , \nonumber\\  
C_{90}^{LMD} & = & c_{53}^{LMD} = 0 \, . \label{C_i_LMD_90}  
\eea
The results for $L_9$ and $L_{10}$ agree with those obtained in
Refs.~\cite{Ecker:1989te,Ecker:1989yg} after employing the two
Weinberg sum rules~\cite{Weinberg_SR} and using, in addition, the
relation $F_V G_V = F_0^2$ which follows from the assumption that the
vector form factor of the pion $F_V^\pi(q^2)$ satisfies an
unsubtracted dispersion relation~\cite{Ecker:1989yg}.

Going through the same steps for the two correlators $\VVP$ and
$\AAP$, we obtain the small momentum expansions 
\bea
\H_V^{LMD}(p^2,q^2,(p+q)^2) &=&
\frac{\langle{\overline\psi}\psi\rangle_0}{(p+q)^2}\,
\Bigg[\,\frac{N_C}{8\pi^2F_0^2} 
\nonumber\\
&&
-\,\frac{p^2+q^2}{M_V^4}\,\bigg(\frac{1}{2}-\frac{N_C}{8\pi^2}\,
\frac{M_V^2}{F_0^2}\,\bigg)  
-\frac{(p+q)^2}{M_V^4}+ \order(p^4) \Bigg] \, ,  \label{H_V_LMD_low}
\nonumber
\\
\H_A^{LMD}(p^2,q^2,(p+q)^2) &=&
\frac{\langle{\overline\psi}\psi\rangle_0}{(p+q)^2}\,
\Bigg[\,\frac{N_C}{24\pi^2F_0^2} 
\nonumber\\
&&
-\,\frac{p^2+q^2}{M_A^4}\,\bigg(\frac{1}{2}-\frac{N_C}{24\pi^2}\,
\frac{M_A^2}{F_0^2}\,\bigg)  
+\frac{(p+q)^2}{M_A^4}+ \order(p^4) \Bigg  ] \, ,  \label{H_A_LMD_low}
\nonumber
\eea
from which we infer the following equations for some of the ${\cal
O}(p^6)$ low-energy constants $A_i$ of Ref.~\cite{Fearing_Scherer} in
the odd intrinsic parity sector, see Eq.~\rf{H_VA_ChPT}, 
\bea
A_2^{LMD} - 4 A_3^{LMD} & = & {F_0^2 \over 8 M_V^4} -
{N_C \over 32 \pi^2} {1\over M_V^2} \, , \nonumber\\
A_2^{LMD} - 2 A_3^{LMD} - 4 A_4^{LMD} & = & - {F_0^2 \over 8 M_V^4} \,
,\nonumber \\   
A_{11}^{LMD} + A_{23}^{LMD} - 4 A_{24}^{LMD} & = & 
{F_0^2 \over 8 M_A^4} - {N_C \over 96 \pi^2} {1 \over M_A^2} \, ,
\nonumber\\ 
3 A_{11}^{LMD} + 4 A_{17}^{LMD} + A_{23}^{LMD} - 2
A_{24}^{LMD} - 4 A_{25}^{LMD} & = & {F_0^2 \over 8 M_A^4}
\, .  \label{A_A_pq2_LMD}
\eea

The numerical values that follow from these expressions for the low-energy 
constants $C_i$ and $A_i$ are discussed in the next section.


\section{Comparison with the resonance Lagrangian approach}
\renewcommand{\theequation}{\arabic{section}.\arabic{equation}}
\setcounter{equation}{0}

In this section, we wish to compare the preceding determination of the
low-energy constants with the approach which uses a Lagrangian with
explicit resonance degrees of freedom. For definiteness, we use the
resonance Lagrangian given in Ref.~\cite{Prades} which has often been
employed in the literature to estimate the low-energy constants at
order $p^6$. In this Lagrangian a vector-field representation is used
for the vector and axial-vector resonances. For convenience, we have
written down this Lagrangian in Appendix~\ref{app:Lag_res}. Note that
no pseudoscalar resonances appear in this Lagrangian.  
Furthermore, as stressed in
Ref.~\cite{Ecker:1989yg}, one has to add local terms from
$\lag_4$,~\footnote{Where $\lag_n = \sum_{k+ 2l = n}
\lag_{(k,l)}$ in terms of the notation introduced in
Eq.~(\ref{lag_kl}).} with fixed coefficients $L_i^{res}$
in order to correctly reproduce the QCD short-distance
behaviour of {\it certain} Green's functions. We shall come back to
this point below.

The calculation of the $\VAP$, $\VVP$ and $\AAP$ three-point functions
with the resonance Lagrangian yields again a result in agreement with
the general solution of the Ward identities with the following
expressions for the invariant functions (the coupling $\beta_V$ is
sometimes denoted by $f_\chi$, see e.g. Ref.~\cite{Bijnens_pipi})
\bea
\F^{res}(p^2,q^2,(p+q)^2) & = & {\langle{\overline\psi}\psi\rangle_0
\over F_0^2 (p+q)^2} \Bigg[ 4 L_9^{res}+  
4 L_{10}^{res} \nonumber \\
&&\qquad + {p^2 \over (p^2 - M_V^2)} \left( f_V^2 - 2 f_V g_V +
2 \sqrt{2} f_V \alpha_V \right) \nonumber \\
&&\qquad + {q^2 \over (q^2 - M_A^2)} \left( - f_A^2 - 2 \sqrt{2} 
f_A \alpha_A \right) \nonumber \\
&&\qquad + {p^2 q^2 \over (p^2 - M_V^2) (q^2 - M_A^2)} \left(
- 2 f_V f_A (A^{(2)} - A^{(3)}) \right) 
\Bigg] \, , \label{F_res} \\
\G^{res}(p^2,q^2,(p+q)^2) & = & {\langle{\overline\psi}\psi\rangle_0
\over F_0^2 q^2 (p+q)^2} \Bigg[ 4 L_9^{res} 
\nonumber \\
&&\qquad + {1\over (p^2 - M_V^2)} \left( - 2 f_V g_V p^2 + 2
\sqrt{2} f_V \alpha_V q^2 - 4 \sqrt{2} f_V \beta_V (p+q)^2 \right)
\nonumber \\ 
&&\qquad + {q^2 \over (q^2 - M_A^2)} \left( - 2 \sqrt{2} f_A
\alpha_A \right) \nonumber \\  
&&\qquad + {(-2 f_V f_A)q^2 \over (p^2 - M_V^2) (q^2 - M_A^2)} \left(
A^{(2)} q^2 - A^{(3)} p^2  + 2 B  (p+q)^2 \right)   
\Bigg] \, ,  \nonumber \\
&& \label{G_res}  
\eea
and 
\bea
\H_V^{ res}(p^2,q^2,(p+q)^2) & = &
- {\langle{\overline\psi}\psi\rangle_0 \over F_0^2 (p+q)^2} \Bigg[ -
{N_C \over 8\pi^2}     
\nonumber \\
&&\qquad + {p^2 \over (p^2 - M_V^2)} \left( 4 \sqrt{2} f_V h_V  \right) 
+ {q^2 \over (q^2 - M_V^2)} \left( 4 \sqrt{2} f_V h_V  \right)
\nonumber \\ 
&&\qquad - {p^2 q^2 \over (p^2 - M_V^2) (q^2 - M_V^2)} \left( 4 f_V^2
\sigma_V \right) \Bigg] \, , \nonumber \\ 
\H_A^{ res}(p^2,q^2,(p+q)^2) & = &
- {\langle{\overline\psi}\psi\rangle_0 \over F_0^2 (p+q)^2} \Bigg[ -
{N_C \over 24\pi^2}     
\nonumber \\
&&\qquad + {p^2 \over (p^2 - M_A^2)} \left( 4 \sqrt{2} f_A h_A  \right) 
+ {q^2 \over (q^2 - M_A^2)} \left( 4 \sqrt{2} f_A h_A  \right)
\nonumber \\ 
&&\qquad - {p^2 q^2 \over (p^2 - M_A^2) (q^2 - M_A^2)} \left( 4 f_A^2
\sigma_A \right) \Bigg] \, . \label{H_VA_res}  
\eea

For momentum transfers that are small as compared to the resonance
masses, we can expand the propagators as sketched in
Eq.~(\ref{expand_prop}). Therefore, the contributions from the
resonance Lagrangian start at ${\cal O}(p^6)$ in the chiral
expansion. The contributions $A^{(2)}, A^{(3)},B$, and $\sigma_V,
\sigma_A$ originate from the exchange of two resonances and start to
contribute to the low-energy expansion at ${\cal O}(p^8)$ only.

\subsection{Even intrinsic parity sector} 

For the $\VAP$ correlator, we thus find the low-energy expansions 
\bea
\F^{res}(p^2,q^2,(p+q)^2) & = & {\langle{\overline\psi}\psi\rangle_0
\over F_0^2 (p+q)^2} \Bigg[ 4 L_9^{res}+  
4 L_{10}^{res} \nonumber \\
&&\qquad - {p^2 \over M_V^2} \left( f_V^2 - 2 f_V g_V +
2 \sqrt{2} f_V \alpha_V \right) \nonumber \\
&&\qquad - {q^2 \over M_A^2} \left( - f_A^2 - 2 \sqrt{2} 
f_A \alpha_A \right) 
+ \order(p^4) \Bigg] \, , \nonumber \\
\G^{res}(p^2,q^2,(p+q)^2) & = & {\langle{\overline\psi}\psi\rangle_0
\over F_0^2 q^2 (p+q)^2} \Bigg[ 4 L_9^{res} 
\nonumber \\
&&\qquad - {1\over M_V^2} \left( - 2 f_V g_V p^2 + 2
\sqrt{2} f_V \alpha_V q^2 - 4 \sqrt{2} f_V \beta_V (p+q)^2 \right)
\nonumber \\ 
&&\qquad - {q^2 \over M_A^2} \left( - 2 \sqrt{2} f_A
\alpha_A \right)
+ \order(p^4) \Bigg] \, . 
 \nonumber
\eea
Comparison with the expressions of the functions $\F$ and $\G$ in
ChPT, Eq.~\rf{FG_ChPT}, leads to the following determination of the
corresponding low-energy constants (again treating $SU(2)$ and $SU(3)$
together)
\bea
C_{78}^{res} & = & c_{44}^{res} = {1\over 4} {1 \over M_V^2}
f_V^2 + {1\over 8} {1 \over M_V^2} f_V g_V + {1\over 2 \sqrt{2}}
{1 \over M_V^2} f_V \beta_V + {1\over \sqrt{2}} {1 \over M_A^2}
f_A \alpha_A \, , \label{C_i_res_78}\nonumber \\
C_{82}^{res} & = & c_{47}^{res} = - {1\over 16} {1 \over
M_V^2} f_V^2 - {1\over 16} {1 \over M_V^2} f_V g_V - {1\over 4
\sqrt{2}} {1 \over M_V^2} f_V \beta_V - {1\over 4 \sqrt{2}} {1
\over M_A^2} f_A \alpha_A \, , \label{C_i_res_82}\nonumber \\
C_{87}^{res} & = & c_{50}^{res} = {1\over 8} {1 \over M_V^2}
f_V^2 - {1\over 8} {1 \over M_A^2} f_A^2 \, ,
\label{C_i_res_87}\nonumber \\ 
C_{88}^{res} & = & c_{51}^{res} = - {1\over 4} {1 \over M_V^2} 
f_V g_V - {1\over \sqrt{2}} {1 \over M_V^2} f_V \beta_V \, ,
\label{C_i_res_88}\nonumber \\  
C_{89}^{res} & = & c_{52}^{res} = {1\over 2} {1 \over M_V^2}
f_V^2 + {1\over 4} {1 \over M_V^2} f_V g_V + {1\over \sqrt{2}}
{1 \over M_V^2} f_V \alpha_V + {1\over \sqrt{2}} {1 \over
M_A^2} f_A \alpha_A \, , \label{C_i_res_89}\nonumber \\  
C_{90}^{res} & = & c_{53}^{res} = - {1\over \sqrt{2}} {1 \over
M_V^2} f_V \beta_V \, .  \label{C_i_res_90} 
\eea
There are two main differences with the LMD ansatz of the
previous section: the absence of a term $(p+q)^2$ in
the low-momentum expansion of 
$\F^{res}(p^2,q^2,(p+q)^2)$, whereas such a term is present in
$\G^{res}(p^2,q^2,(p+q)^2)$ but not in $\G^{LMD}(p^2,q^2,(p+q)^2)$.

In Table~\ref{tab:C_i_LMD_res} (see e.g. Eq.~(\ref{C_i_LMD_90}) for
the translation into the corresponding $SU(2)$ constants $c_i$) we
compare the numerical values for the low-energy constants $C_i$ in the
LMD approximation with those obtained from the resonance Lagrangian.
As recalled in the introduction and as discussed in
Ref.~\cite{Ecker:1989te}, these numbers have to be understood as the
values of the low-energy constants at the scale set by $\Lambda_H$,
which we identify with the $\rho$ mass $M_V$.  We have used the values
$F_0 = 92.4$~MeV, $M_V = 769$~MeV, and $M_A = 1230$~MeV, as well as
the two sets of input values for the resonance parameters listed in
Table~\ref{tab:res_param} as given in Ref.~\cite{Prades} (Set I),
based on an ENJL model, and from
Refs.~\cite{Bijnens_pipi,ABT_res_param} (Set II), extracted from
resonance decays.  If we allow for a relative error of about 30\% in
these values, a typical size for the uncertainty attached to a
large-$N_C$ estimate, the agreement is rather good, except for the
cases of $C_{88}$ and $C_{90}$.  We postpone the discussion of some
phenomenological implications of the differences shown by
Table~\ref{tab:C_i_LMD_res} to section 8 below.

\begin{table}[h]
\caption{Numerical values for the low-energy constants $C_i$ (in
units of $10^{-4} / F_0^2$) obtained from the LMD estimates in
Eq.~(\protect\ref{C_i_LMD_90}) and with the expressions derived from the
resonance Lagrangian in Eq.~(\protect\ref{C_i_res_90}) for two
different sets of input values for the resonance parameters.}
\begin{center}
\renewcommand{\arraystretch}{1.1}
\begin{tabular}{|l|r@{.}l|r@{.}l|r@{.}l|r@{.}l|r@{.}l|r@{.}l|}
\hline
 & \multicolumn{2}{|c|}{{$C_{78}$}}  
 & \multicolumn{2}{|c|}{{$C_{82}$}}  
 & \multicolumn{2}{|c|}{{$C_{87}$}}  
 & \multicolumn{2}{|c|}{{$C_{88}$}}  
 & \multicolumn{2}{|c|}{{$C_{89}$}}  
 & \multicolumn{2}{|c|}{{$C_{90}$}}  
\\ 
\hline  
LMD    & 1 & 09 & -0 & 36 & 0 & 40 & -0 & 52 & 1 & 97 
& 0 & 0  \\ 
Set I  & 1 & 09 & -0 & 29 & 0 & 47 & -0 & 16 & 2 & 29
& 0 & 33 \\
Set II & 1 & 49 & -0 & 39 & 0 & 65 & -0 & 14 & 3 & 22
&  0 & 51 \\
\hline
\end{tabular}
\label{tab:C_i_LMD_res}  
\end{center}
\end{table}

\begin{table}[h]
\caption{Values for the parameters in the resonance Lagrangian from
Ref.~\protect\cite{Prades} (Set I) and from
Refs.~\protect\cite{Bijnens_pipi,ABT_res_param} (Set II).}
\begin{center}
\renewcommand{\arraystretch}{1.1}
\begin{tabular}{|l|r@{.}l|r@{.}l|r@{.}l|r@{.}l|r@{.}l|r@{.}l|}
\hline 
 & \multicolumn{2}{|c|}{{$f_V$}} & \multicolumn{2}{|c|}{{$g_V$}} 
 & \multicolumn{2}{|c|}{{$\alpha_V$}} 
 & \multicolumn{2}{|c|}{{$\beta_V \equiv f_\chi$}} 
 & \multicolumn{2}{|c|}{{$f_A$}} & \multicolumn{2}{|c|}{{$\alpha_A$}}  
\\ \hline  
Set I
& 0 & 17 & 0 & 08 & -0 & 015 &\hspace*{1.25mm} -0 & 019 & 0 & 085 & -0
& 0092 
\\
Set II 
& 0 & 20 & 0 & 09 & -0 & 014 & -0 & 025 & 0 & 10 & -0 & 0067
\\
\hline
\end{tabular}
\label{tab:res_param}  
\end{center}
\end{table}

The values displayed in the second line of Table 1 were obtained by taking 
$b=2M_A^2$ in the LMD ansatz (\ref{FG_LMD}). If we had taken $b=4M_A^2/3$ 
instead, these values would have changed within an 
acceptable range: about 20\% for $C_{78}$ and $C_{82}$, 16\% in the case of 
$C_{89}$. The biggest variation occurs for $C_{88}$, around 30\%, while 
$C_{87}$ and $C_{90}$ remain unchanged, being insensitive to the value of $b$. 
Notice also that $L_9$ is proportional to $b$, $L_9=F_0^2b/4M_V^2M_A^2$. 
Changing $b$ from $2M_A^2$ to $4M_A^2/3$ decreases the value of $L_9$ by as 
much as 33\%, while leaving $L_{10}$ unaffected.
This modifies the ${\cal O}(p^4)$ prediction of the pion charge 
radius from 
$\langle r^2\rangle^{\pi}_V(b=2M_A^2)=0.47\pm 0.13$ fm$^2$ to 
$\langle r^2 \rangle^{\pi}_V(b=4M_A^2/3)=0.33\pm 0.09$ fm$^2$, 
somewhat lower than the experimental value 
$\langle r^2 \rangle^{\pi}_V=0.439\pm 0.008$ fm$^2$ 
\cite{Amendolia}.

At this stage, it is worthwhile to stress that it is not possible to find a
one-to-one correspondence between the parameter sets of the resonance 
Lagrangian and of the LMD approximation to large-$N_C$ QCD. 
This directly reflects the differences mentioned above in
the low-energy expansions of the functions $\F^{LMD}$ and $\G^{LMD}$
on the one hand, and of $\F^{res}$ and $\G^{res}$ on the other
hand. Nevertheless, some agreement can be obtained by adjusting the
resonance parameters.  When performing such a parametric comparison,
one has to observe that the LMD ansatz already encodes some additional
information. For instance, the two Weinberg sum
rules~\cite{Weinberg_SR} are fulfilled. In the LMD approximation they
take the form $F_V^2 = F_0^2 + F_A^2$ and $F_V^2 M_V^2 = F_A^2 M_A^2$
and allow one to express $F_V$ and $F_A$ through the resonance masses
and $F_0$.  Furthermore, in the LMD approximation the identity $F_V
G_V = F_0^2$ holds. Finally, one has to make the following
identifications between the parameters in the vector- and the
tensor-field representation for the resonances: $f_V \equiv F_V / M_V,
g_V \equiv G_V / M_V$ and $f_A \equiv F_A / M_A$, see
Ref.~\cite{Ecker:1989yg}.

Comparing the expressions for $C_{90}$ and $C_{88}$ in the LMD
approximation, Eq.~(\ref{C_i_LMD_90}), with those obtained from the
resonance Lagrangian, Eq.~(\ref{C_i_res_90}), and using $F_V G_V =
F_0^2$ we get
\bea
\beta_V = 0 \, . 
\nonumber
\eea
This removes the largest numerical discrepancies in
Table~\ref{tab:C_i_LMD_res}. Note in particular the huge cancellation
in $C_{88}^{res}$ for the values of $\beta_V$ given in
Table~\ref{tab:res_param}.  On the other hand, using the Weinberg sum
rules and the identifications mentioned above one notices that
\bea
C_{87}^{LMD} \equiv C_{87}^{res} \, .  
\nonumber
\eea
Thus we are left with the three equations $C_i^{LMD} = C_i^{res},
i = 78, 82, 89,$ for the remaining two unknowns $\alpha_V$ and
$\alpha_A$. It turns out that  this system of equations is
inconsistent. We can solve, however, for $\alpha_V$, since in the
difference $C_{78} - C_{89}$ the term with $f_A \alpha_A$ drops
out. From the requirement $C_{78}^{LMD} - C_{89}^{LMD} =
C_{78}^{res} - C_{89}^{res}$ we obtain 
\bea
\alpha_V = - {\sqrt{2} \over 8} {F_0 M_V \over M_A^3} {(M_A^2 +
M_V^2) \over \sqrt{M_A^2 - M_V^2}} 
 = - 0.015  \, ,  
\nonumber
\eea
in remarkable agreement with the values quoted in
Refs.~\cite{Prades,Bijnens_pipi,ABT_res_param}, see
Table~\ref{tab:res_param}.
On the other hand, requiring $C_{78}^{LMD} + 4 C_{82}^{LMD} =
C_{78}^{res} + 4 C_{82}^{res}$, leads to 
\bea
- {1\over 8} {F_0^2 \over M_V^2 M_A^2} = 0 \, .
\nonumber
\eea
This is not compatible with the spontaneous breakdown of chiral
symmetry~\cite{tHooft} and the Goldstone theorem, 
which requires $F_0 \neq 0$.

\subsection{Odd intrinsic parity sector} 

The low-energy expansion of the resonance expressions for the two
correlators $\VVP$ and $\AAP$ gives the following estimates of the
low-energy constants $A_i$ 
\bea
A_2^{res} - 4 A_3^{res} & = & - \sqrt{2} f_V h_V {1 \over M_V^2}
\, , \label{A_V_p2_res} \nonumber \\
A_2^{res} - 2 A_3^{res} - 4 A_4^{res} & = & 0 \, , \nonumber \\ 
A_{11}^{res} + A_{23}^{res} - 4 A_{24}^{res} & = & - \sqrt{2}
f_A h_A {1 \over M_A^2} \, , \nonumber \\ 
3 A_{11}^{res} + 4 A_{17}^{res} + A_{23}^{res} - 2
A_{24}^{res} - 4 A_{25}^{res} & = & 0 \, . \label{A_A_pq2_res}
\eea

In Table~\ref{tab:A_i_LMD_res} we compare the numerical values for the
low-energy constants $A_i$ from the LMD estimates in
Eq.~(\ref{A_A_pq2_LMD}) with those from the resonance Lagrangian,
Eq.~(\ref{A_A_pq2_res}). We have introduced the following notations
for the combinations of low-energy constants that appear in $\VVP$ and
$\AAP$
\bea
A_{V,p^2} & = & A_2 - 4 A_3 \, ,  \nonumber \\ 
A_{V,(p+q)^2} & = & A_2 - 2 A_3 - 4 A_4 \, , \nonumber \\  
A_{A,p^2} & = & A_{11} + A_{23} - 4 A_{24} \, , 
\nonumber \\ 
A_{A,(p+q)^2} & = & 3 A_{11} + 4 A_{17} + A_{23} - 2 A_{24} - 4
A_{25}\, . \label{A_combinations} 
\eea
\begin{table}[h]
\caption{Numerical values for the combinations of low-energy constants 
$A_i$ defined in Eq.~(\protect\ref{A_combinations}), in units of
$10^{-4} / F_0^2$, obtained from the LMD estimates in
Eq.~(\protect\ref{A_A_pq2_LMD}) and from the resonance Lagrangian in
Eq.~(\protect\ref{A_A_pq2_res}) (Set I).} 
\begin{center}
\renewcommand{\arraystretch}{1.1}
\begin{tabular}{|l|r@{.}l|r@{.}l|r@{.}l|r@{.}l|}
\hline  
 & \multicolumn{2}{|c|}{{$A_{V,p^2}$}}  
 & \multicolumn{2}{|c|}{{$A_{V,(p+q)^2}$}}  
 & \multicolumn{2}{|c|}{{$A_{A,p^2}$}}  
 & \multicolumn{2}{|c|}{{$A_{A,(p+q)^2}$}}  
\\ \hline  
LMD   & -1 & 11 &\hspace*{2mm} -0 & 26 & -0 & 14 &\hspace*{2mm} 0 & 040 \\
Set I & -1 & 13 &   0 & 0 & -0 & 096 &  0 &  0 \\
\hline
\end{tabular}
\label{tab:A_i_LMD_res}
\end{center}
\end{table}

The agreement for the low-energy constants $A_{V,A,p^2}$ is quite
good, whereas the two approaches give different results for
$A_{V,A,(p+q)^2}$. In the expressions for the low-energy constants
$A_i$ from the resonance Lagrangian (\ref{A_A_pq2_res}) we used the
ENJL estimates (Set I)
\bea
f_V h_V & = & {N_C \over 16 \pi^2} {\sqrt{2} \over 8} (1 + g_A) =
0.0055 \, , \nonumber \\
f_A h_A & = & {N_C \over 16 \pi^2} {\sqrt{2} \over 24} g_A (1 + g_A) = 
0.0012 \, , \label{fVhV_fAhA_ENJL}
\eea
with $N_C = 3$ and $g_A = 0.65$, see Ref.~\cite{Prades}, in particular
the Erratum. We shall discuss estimates for the low-energy
constants $A_{V,p^2}$ and $A_{V,(p+q)^2}$ beyond the LMD approximation 
in section~\ref{sec:pigammagamma}.

Again, it is impossible to find a one-to-one relation between the
parameters of the resonance Lagrangian and those that describe the LMD
ansatz. The reason lies in the absence of a term proportional to
$(p+q)^2$ in the low-energy expansions of both $\H_V^{res}$ and
$\H_A^{res}$. We notice nevertheless that requiring $A_{V,p^2}^{LMD} =
A_{V,p^2}^{res}$ and $A_{A,p^2}^{LMD} = A_{A,p^2}^{res}$
leads to the relations 
\bea
f_V h_V & = & {1\over \sqrt{2}} \left( {N_C \over 32 \pi^2} - {F_0^2
\over 8 M_V^2} \right) = 0.0054 \, , \nonumber \\ 
f_A h_A & = & {1\over \sqrt{2}} \left( {N_C \over 96 \pi^2} - {F_0^2
\over 8 M_A^2} \right) = 0.0017 \, , \nonumber 
\eea
in rather good agreement with Eq.~(\ref{fVhV_fAhA_ENJL}).  

We conclude that, although an adequate adjustment of the parameters of
the resonance Lagrangian of Ref.~\cite{Prades} can bring the
determinations of the low-energy constants considered here to a
reasonable numerical agreement, there is no way to establish an
algebraic equivalence between the two approaches.  We have focused
here on the Lagrangian of Ref.~\cite{Prades} which is the most
complete as far as couplings among resonances and to the Goldstone
bosons are concerned. Therefore similar conclusions will also hold for
other existing Lagrangians with resonance fields, see
Refs.~\cite{Gasio69,Meissner:1988ge,Bando88,Donoghue:1989ed,Ecker:1989te,Birse:1996hd,Ecker:1989yg}
and references therein.


\section{QCD short-distance constraints on the resonance Lagrangian}
\renewcommand{\theequation}{\arabic{section}.\arabic{equation}}
\setcounter{equation}{0}

The discrepancies between the estimates for the low-energy constants
at ${\cal O}(p^6)$ from the LMD ansatz and from the resonance
Lagrangian, as observed in the previous section, can be traced back to
the different high-energy behaviours of the corresponding Green's
functions in the two approaches. In fact, as we shall show in this
section, the Green's function derived from the resonance Lagrangian
are incompatible with the QCD short-distance constraints. 

\subsection{Two large momenta}

We first consider the limit when the two momenta become simultaneously 
large, see
Eqs.~\rf{F_OPE} -- \rf{H_OPE}. The invariant function $\F^{res}$ from
Eq.~(\ref{F_res}) behaves as
\bea
\lefteqn{\lim_{\lambda \to \infty} \F^{res}((\lambda
p)^2, (\lambda q)^2, (\lambda p + \lambda q)^2) =
{\langle{\overline\psi}\psi\rangle_0 \over \lambda^2 F_0^2 (p+q)^2} 
\Big[ 4 L_9 + 4 L_{10} + r_{FV} + r_{FA} + r_{FV\!A} \Big] } 
\nonumber \hspace{4cm} \\ 
&&\!\!\!\!\!\!\!
 + {\langle{\overline\psi}\psi\rangle_0 \over \lambda^4 F_0^2} \left[
{(r_{FV} + r_{FV\!A}) M_V^2 \over p^2 (p+q)^2} + {(r_{FA} + r_{FV\!A})
M_A^2 \over q^2 (p+q)^2} \right] + \order({1\over \lambda^6}) \, ,
\label{OPE_F_res}  
\eea
where $r_{FV}, r_{FA},$ and $r_{FV\!A}$ denote the coefficients of the
terms $p^2/(p^2 - M_V^2), q^2/(q^2 - M_A^2),$ and $(p^2 q^2) / ((p^2 - 
M_V^2) (q^2 - M_A^2))$ in Eq.~(\ref{F_res}), respectively. 
Agreement with the OPE result from Eq.~\rf{F_OPE} at order
$1 / \lambda^2$ can be achieved if the constraint
\be \label{constr_F1}
4 L_9 + 4 L_{10} + f_V^2 - 2 f_V g_V + 2 \sqrt{2} f_V \alpha_V -
f_A^2 - 2 \sqrt{2} f_A \alpha_A - 2 f_V f_A (A^{(2)} - A^{(3)}) = 0 \,  
\ee
is fulfilled. However, at order $1 / \lambda^4$, we observe that
$\F^{res}$ is not compatible with the OPE result in
Eq.~\rf{F_OPE}, since the term $\sim 1 / (p^2 q^2)$ is missing in
Eq.~(\ref{OPE_F_res}).  

As was shown in Ref.~\cite{Ecker:1989yg}, requiring agreement of
certain Green's functions with the short-distance properties of QCD
uniquely determines the low-energy constants $L_i$ at ${\cal O}(p^4)$,
even though the contributions of the resonance Lagrangian in the
vector-field representation only start at ${\cal O}(p^6)$.\footnote{If
a tensor-field formulation is used for the vector and axial-vector
resonances, the contributions from the resonances start already at
order $p^4$ and lead directly to the usual estimates for the
$L_{i}$. On the other hand, several couplings which appear in the
resonance Lagrangian of Ref.~\cite{Prades}, like $\alpha_V,\beta_V$ or
$\alpha_A$, cannot be written down in the tensor-field representation,
at least not without introducing additional derivatives.}  In fact,
one has to add local terms from $\lag_4$ to the resonance Lagrangian
in order to obtain the correct short-distance behaviour of these
Green's functions. Analogously, at ${\cal O}(p^6)$, one might try to
add local counterterms from $\lag_6$. This is, however, not enough to
bring the function $\F^{res}$ in agreement with the OPE. One has as
well to add new terms involving both resonance fields and additional
derivatives. A similar observation, concerning the $\langle V V
S\rangle$ correlator, was made in Ref.~\cite{Moussallam:1994at}: the
short-distance behaviour of the three-point function $\langle V V
S\rangle$ becomes consistent with the OPE only if a term $\langle
(\hat V_{\mu\nu} - (f_V /2) f_{\mu\nu}^+)^2 \nabla^2 \hat S \rangle$
is added to the resonance Lagrangian. In the present case the
situation is more involved, i.e.\ several new terms would have to be
added. We have not undertaken the task to construct
them explicitly. One can show, however, that if one matches $\F^{res}$
with the constraints imposed by the OPE, the local counterterms at
${\cal O}(p^6)$ have to be adjusted in such a way that one finally
obtains the same values for the low-energy constants as with the LMD
approach. We caution the reader that the mere fact of using a tensor field
representation for the resonances does not, by itself, guarantee to
yield the same estimates for the low-energy constants as with the LMD
ansatz. Actually, it was pointed out in Ref.~\cite{Moussallam97} that the
resonance Lagrangian with a tensor field representation leads to a
correlator $\VAP$~\cite{Baur_Urech} which does not have the correct
short-distance properties.

We may perform a similar analysis for $\G^{res}$ from
Eq.~(\ref{G_res}). In this case, one may obtain agreement with the QCD
result of Eq.~\rf{G_OPE}, provided the coupling constants in the
resonance Lagrangian are adjusted as follows
\bea
& & f_V g_V M_V^2 + \sqrt{2} f_V \alpha_V M_A^2 - \sqrt{2} f_A \alpha_A
M_V^2 =  {F_0^2 \over 2} \, , \quad L_9 = {1\over 2} f_V g_V \, ,
\nonumber \\
&& A^{(2)} = \sqrt{2}\, {\alpha_V \over f_A} \, , \quad A^{(3)} =
\sqrt{2}\, {\alpha_A \over f_V} \, , \quad \beta_V = 0 \, , \quad 
B = 0 \, .  \label{constr_G}  
\eea
We note, however, that in deriving these constraints we have not taken
into account possible contributions to $\G^{res}$ from the new local terms
that have to be added to the resonance Lagrangian in order to make
$\F^{res}$ compatible with the OPE. The expression for $L_9$ in 
Eq.~(\ref{constr_G}) agrees
with Refs.~\cite{Ecker:1989te,Ecker:1989yg}, whereas the relations for
$A^{(2)}$ and $A^{(3)}$ agree with those given in Ref.~\cite{Prades}.

The short-distance behaviour of the functions $\H_V^{res}$ and
$\H_A^{res}$ in Eq.~(\ref{H_VA_res}) in the odd intrinsic parity
sector is given by
\bea
\lim_{\lambda \to \infty} \H_V^{res}((\lambda
p)^2, (\lambda q)^2, (\lambda p + \lambda q)^2) &\!\!= &
\!\!-{\langle{\overline\psi}\psi\rangle_0 \over \lambda^2 F_0^2 (p+q)^2}
\left[ - {N_C \over 8\pi^2} + 8 \sqrt{2} f_V h_V - 4 f_V^2 \sigma_V
\right] \nonumber \\ 
&&\!\!-{\langle{\overline\psi}\psi\rangle_0 M_V^2 \over \lambda^4 F_0^2}
\left[ {4 \sqrt{2} f_V h_V - 4 f_V^2 \sigma_V \over p^2 (p+q)^2} + {4
\sqrt{2} f_V h_V - 4 f_V^2 \sigma_V \over q^2 (p+q)^2} \right]
\nonumber \\   
&&\!\!+ \order({1\over \lambda^6}) \, , \nonumber \\ 
\lim_{\lambda \to \infty} \H_A^{res}((\lambda
p)^2, (\lambda q)^2, (\lambda p + \lambda q)^2) &\!\!= &
\!\!-{\langle{\overline\psi}\psi\rangle_0 \over \lambda^2 F_0^2 (p+q)^2}
\left[ - {N_C \over 24\pi^2} + 8 \sqrt{2} f_A h_A - 4 f_A^2 \sigma_A
\right] \nonumber \\ 
&&\!\!- {\langle{\overline\psi}\psi\rangle_0 M_A^2 \over \lambda^4 F_0^2}
\left[ {4 \sqrt{2} f_A h_A - 4 f_A^2 \sigma_A \over p^2 (p+q)^2} + {4
\sqrt{2} f_A h_A - 4 f_A^2 \sigma_A \over q^2 (p+q)^2} \right]
\nonumber \\   
&&\!\!+ \order({1\over \lambda^6}) \, .  
\label{OPE_H_VA_res} 
\eea
Comparison of these expressions with the OPE result of Eq.~\rf{H_OPE}
leads, at order $1 / \lambda^2$, to the constraints
\bea 
- {N_C \over 8\pi^2} + 8 \sqrt{2} f_V h_V  - 4 f_V^2 \sigma_V & = & 0 , 
\nonumber \\ 
- {N_C \over 24\pi^2} + 8 \sqrt{2} f_A h_A  - 4 f_A^2 \sigma_A & = & 0
. \label{constr_H_VA_1}  
\eea
 
However, at order $1 / \lambda^4$, we again observe that
$\H_{V}^{res}$ and $\H_{A}^{res}$ are not consistent with the OPE
result, since the terms $\sim 1 / (p^2 q^2)$ are missing in
Eq.~(\ref{OPE_H_VA_res}).

\subsection{One large momentum}

In section~\ref{sec:short_distance} we have also derived the
short-distance properties of the Green's functions $\VAP, \VVP$ and
$\AAP$ when the relative distance between only two of the currents
becomes small. In certain physical applications only this limit is
relevant and we shall now discuss the corresponding constraints on the
parameters in the resonance Lagrangian. We note, however, that it is
not possible to satisfy simultaneously {\it all} the constraints given
below. Moreover, some of these constraints are in contradiction with
the relations derived in the previous section. These inconsistencies
can again be traced back to the fact that the Green's functions
derived from the resonance Lagrangian do not correctly reproduce the
QCD short-distance behaviour.

The constraints on the functions $\F$ and $\G$ from Eqs.~\rf{VAPope2}
and \rf{VAPope3}, when the space-time arguments of the vector and
axial-vector currents coincide, can be satisfied provided the
resonance parameters obey the relations
\be \label{relations_1} 
A^{(2)} = \sqrt{2}\, {\alpha_V \over f_A}, \quad 
B = - \sqrt{2}\, {\beta_V \over f_A} \, . 
\ee
Furthermore, we recover the usual resonance estimate $L_{10} = - f_V^2
/ 4 + f_A^2 / 4$ and the first Weinberg sum rule $M_V^2 f_V^2 = F_0^2
+ M_A^2 f_A^2$.  Imposing the relation $L_9 = f_V g_V / 2$, we
furthermore get the relations
\be \label{relations_2} 
A^{(3)} = \sqrt{2} {\alpha_A \over f_V} , \quad 
M_V^2 f_V^2 - M_V^2 f_V g_V + \sqrt{2} M_A^2 f_V \alpha_V + \sqrt{2} 
M_V^2 f_A \alpha_A = 0 \, . 
\ee

The second limit, when the distance between the vector current
and the pseudoscalar density becomes small, corresponding to the
constraints~\rf{VAPope4}, can be satisfied provided
\be
A^{(3)} = \sqrt{2}\, {\alpha_A \over f_V} , \quad 
\beta_V = 0, \quad 
B = 0 \, . 
\ee
We also get the result $L_9 = f_V g_V / 2$. Imposing the usual
resonance estimate for $L_{10}$, we obtain the additional relations
\be
\sqrt{2} f_V \alpha_V = - {f_A^2 \over 2}, \quad 
A^{(2)} = \sqrt{2}\, {\alpha_V \over f_A} \, . 
\ee

Finally, the constraints~\rf{VAPope5}, when the arguments of the
axial-vector current and the pseudoscalar density coincide, lead again
to the relations~(\ref{relations_1}). Using in addition the usual
resonance estimates for $L_9$ and $L_{10}$, we obtain
\be \label{relations_3} 
A^{(3)} = \sqrt{2}\, {\alpha_A \over f_V} , \quad 
f_V g_V - {f_V^2 \over 2} - \sqrt{2} f_A \alpha_A = {F_0^2 \over 2
M_V^2} \, . 
\ee
We note that we recover in the first case, corresponding to
Eqs.~\rf{VAPope2} and \rf{VAPope3}, all the constraints from the
leading terms in the expansion in $1 / \lambda$, when all momenta in
$\F$ and $\G$ become large. In the latter two cases we get, however,
only a subset of these relations. 
 
The equations (\ref{relations_1}) -- (\ref{relations_3}) arising from
the three different limits are perfectly compatible.  If however we
combine them with the first relation in Eq.~(\ref{constr_G}) from the
OPE constraint when all momenta in $\G$ become large, we find a
contradiction.

We now turn to the functions $\VVP$ and $\AAP$. We first consider the
case when the two vector currents are taken at the same point, which
is relevant, for instance, in the decay $P
\to l^+ l^-$, where $P = \pi^0, \eta$ (see the discussion in
Ref.~\cite{Pi_ll}) or when the space-time arguments of the two
axial-vector currents in $\AAP$ coincide. The corresponding constraint
(\ref{H_V/A_OPE}) can be satisfied, provided the resonance parameters
fulfill the relations
\be \label{H_V_A_res_x_to_y}
\sqrt{2} f_X h_X = c_X {N_C \over 32 \pi^2} - {1\over 8} {F_0^2 \over
M_X^2} , \quad 
f_X^2 \sigma_X = c_X {N_C \over 32 \pi^2} - {1\over 4} {F_0^2 \over
M_X^2} , \quad X = V,A, 
\ee 
with $c_V = 1, c_A = 1/3$.  In the case when the space-time argument
of one of the vector currents and of the pseudoscalar density in
$\VVP$ coincide, the constraint from Eq.~(\ref{H_V_x_to_0}) can be
satisfied, if
\be \label{H_V_res_x_to_0}
\sqrt{2} f_V h_V = {N_C \over 32 \pi^2} - {1\over 4} {F_0^2 \over
M_V^2} , \quad 
f_V^2 \sigma_V = {N_C \over 32 \pi^2} - {1\over 2} {F_0^2 \over
M_V^2} , 
\ee
where we have also used the short-distance constraint (\ref{VT_OPE})
on $\Pi_{VT}(p^2)$. The same limit in $\AAP$, see
Eq.~(\ref{H_A_x_to_0}), yields the relations
\be \label{H_A_res_x_to_0} 
\sqrt{2} f_A h_A = {N_C \over 96 \pi^2} , \quad
f_A^2 \sigma_A = {N_C \over 96 \pi^2} \, . 
\ee
We recover in all cases the constraints (\ref{constr_H_VA_1}) when all
momenta in $\H_{V,A}$ become large. We note, however, that although
one can individually satisfy the constraints from the OPE, the
relations (\ref{H_V_res_x_to_0}) and (\ref{H_A_res_x_to_0}) are
incompatible with Eq.~(\ref{H_V_A_res_x_to_y}).

\newpage
\section{Resonance contributions to pion form factors}
\renewcommand{\theequation}{\arabic{section}.\arabic{equation}}
\setcounter{equation}{0}

 In this section, we discuss a few phenomenological applications of the 
various ans\"atze considered previously. The first two examples, the vector
form factor of the pion and the radiative pion decay, involve the low-energy 
constants $C_i$ that were determined in section 5. The third example, the 
pion-photon-photon transition form factor, illustrates a situation where the 
MHA goes beyond the simplest LMD approximation.

\subsection{Vector form factor of the pion}

There are two combinations of renormalized low-energy constants
$c_i^r$ from the chiral Lagrangian $\lag_6$ that enter in the vector
form factor of the pion, see Ref.~\cite{Renorm_p6}
\bea\
r_{V1}^r & = & -16 c_6^r - 4 c_{35}^r - 8 c_{53}^r \, , \nonumber \\
r_{V2}^r & = & - 4 c_{51}^r + 4 c_{53}^r \, . \nonumber 
\eea
The resonance estimates for $r_{V1}^r$ and $r_{V2}^r$ given in
Ref.~\cite{BCT} read (recall that $f_\chi \equiv \beta_V$) 
\bea
r_{V1}^{r,res}(\mu = M_V) & = & {2 \sqrt{2} f_\chi f_V F^2 \over
M_V^2} \, , \nonumber \\
r_{V2}^{r,res}(\mu = M_V) & = & {g_V f_V F^2 \over M_V^2} \, .
\label{rV2_BCT} 
\eea
Using the resonance estimates for the constants $c_i^{res}$ given in
Eqs.~(\ref{C_i_res_90}) we obtain the same result for
$r_{V2}^{r,res}$, if we identify the pion decay constant $F$ with
$F_0$.  

With the LMD estimates from Eqs.~(\ref{C_i_LMD_90}) we get
\bea
r_{V2}^{r,LMD}(\mu = M_V) = {F_0^4 \over M_V^4} \, , 
\eea
which agrees with the result given in Eq.~(\ref{rV2_BCT}), if we use
the relation $F_V G_V = F_0^2$ which is valid within the LMD
ansatz. From the present analysis, we cannot obtain an estimate for
$r_{V1}^r$, which describes the quark mass corrections to the value of
the vector form factor at vanishing momentum transfer, see
Ref.~\cite{BCT}.

\subsection{The decay $\pi \to e \nu_e \gamma$}

The decay $\pi(p) \to e \nu_e \gamma(q)$ is described by two form
factors $V$ and $A$. The contribution from $\lag_6$ to $A$ can be
written as~\cite{BT,Renorm_p6}
\bea
A((p-q)^2) = M_\pi^2 r_{A1}^r + (p \cdot q) r_{A2}^r \ , \nonumber 
\eea
with 
\bea
r_{A1}^r & = & 48 c_6^r - 16 c_{34}^r + 8 c_{35}^r - 8 c_{44}^r + 16
c_{46}^r - 16 c_{47}^r + 8 c_{50}^r \, , \nonumber \\
r_{A2}^r & = & 8 c_{44}^r - 16 c_{50}^r + 4 c_{51}^r \, . \nonumber 
\eea

With the estimates for the low-energy constants $c_i^r$ obtained from
the resonance Lagrangian, Eqs.~(\ref{C_i_res_90}), we obtain
\be
r_{A2}^{r,res}(\mu = M_V) = F^2 {2 \over M_A^2} \left( f_A^2 + f_A
\alpha_A 2 \sqrt{2} \right) \approx 0.55 \cdot 10^{-4} \, ,
\label{rA2_res}   
\ee 
where we used the same values $f_A = 0.080, \alpha_A = - 6.66
\cdot 10^{-3}$, as in Ref.~\cite{BT} and $F \approx F_0 = 92.4$~MeV, and
$M_A = 1230$~MeV.  

On the other hand, using the LMD estimates given in
Eqs~(\ref{C_i_LMD_90}) we obtain
\bea
r_{A2}^{r,LMD}(\mu = M_V) = {F_0^4 \over M_V^2 M_A^2} - 2 {F_0^4 \over
M_A^4} \approx 0.18 \cdot 10^{-4} \, . 
\eea
The LMD estimate differs by a factor three from the value given in
Eq.~(\ref{rA2_res}). In fact, since we obtain a different relative
sign in our result for $r_{A2}^{r,res}$ as compared to the expression
given in Ref.~\cite{BT}, the LMD estimate is even a factor of five
smaller than the value $r_{A2}^{r,res}(\mu = M_V) \approx 0.89 \cdot
10^{-4}$ used in that paper.

Since the same combination of resonance parameters that determines
$r_{A2}^r$ also enters in the decay amplitude for $a_1 \to \pi
\gamma$, the LMD ansatz predicts a decay rate more than an order of
magnitude smaller than the usual value \cite{PDG2000}. However, the
experimental situation concerning this decay is far from being
clear~\cite{a1_to_pigamma}, see the remarks following Eq.~(62) in
Ref.~\cite{Moussallam97}.

\subsection{Pion-photon-photon transition form factor}
\label{sec:pigammagamma}

Up to now, we have shown that a very minimal ansatz allows to take
into account the {\it leading} asymptotic behaviours of the three
point functions $\VAP$, $\VVP$ and $\AAP$. This simple representation
would in general certainly not be sufficient if additional information
were added. However, within the large-$N_C$ framework considered here,
one always has the freedom to go beyond the LMD representation and to
add other zero-width resonance states. 
In this section, we wish to illustrate this point by
considering the form factor $\F_{\pi\gamma^*\gamma^*}(q_1^2,q_2^2)$
that describes the transition between a pion and two (possibly
off-shell) photons in the chiral limit. This form factor is defined as
\bea
i\int d^4x\,e^{iq\cdot x}
\,\langle 0 \vert T\{j_{\mu}(x)j_{\nu}(0)\}\vert \pi^0(p) \rangle = 
\epsilon_{\mu\nu\alpha\beta}q^{\alpha}p^{\beta}\,
\F_{\pi\gamma^*\gamma^*}(q^2,(p-q)^2) 
\eea
or still
\bea
\F_{\pi\gamma^*\gamma^*}(q_1^2,q_2^2) =
  - {2\over 3} {F_0 \over \qbarq} \lim_{(q_1+q_2)^2 \to 0}  (q_1 +
q_2)^2 \H_V(q_1^2, q_2^2, (q_1+q_2)^2) \, .  \nonumber 
\eea
{F}or both photons on-shell, the value
of this form factor is fixed by the Wess-Zumino-Witten anomaly term,
\bea
\F_{\pi\gamma^*\gamma^*}(0,0) = -\,{N_C \over 12 \pi^2 F_0} \,
. \nonumber 
\eea
Many studies have been devoted to this form factor in the past, see  
\cite{Brodsky_Lepage,formfactor} and references therein. 
In particular, its behaviour in the limit $Q^2\to\infty$, $\omega$ 
fixed, with $-Q^2=(q_1^2+q^2_2)$, $\omega = (q_1^2-q_2^2)/(q_1^2+q_2^2)$, 
has been investigated, with the result
\bea
\F_{\pi\gamma^*\gamma^*}(q_1^2,q_2^2) =
-\, \frac{4F_0}{3}\,\frac{f(\omega)}{Q^2} + {\cal
O}\left(\frac{1}{Q^4}\right) \, , 
\eea
for $-1<\omega <1$. The function $f(\omega)$ can be expressed in terms of 
the pion distribution function $\varphi_{\pi}(u)$ 
\cite{Brodsky_Lepage}   
\bea
f(\omega) = \int_0^1 du\,
\frac{\varphi_{\pi}(u)}{(1-u)(1+\omega)+u(1-\omega)}
\,\big[ 1 + {\cal O}(\alpha_s) \big] \, , 
\nonumber
\eea
normalized as 
\bea
\int_0^1 du \,\varphi_{\pi}(u) = 1 \, . 
\nonumber
\eea
This last condition is sufficient in order to study the limit 
$-q^2_1=-q_2^2=Q^2/2\to\infty$ and leads to the result~\cite{novikov}
\bea
\F_{\pi\gamma^*\gamma^*}(-\,\frac{Q^2}{2},-\frac{Q^2}{2}) = 
-\frac{4F_0}{3}\,\frac{1}{Q^2}
 + {\cal O}\left(\frac{1}{Q^4}\right) \,.
\nonumber
\eea
The function $\varphi_{\pi}(u)$ is only known asymptotically, and this
asymptotic expression is reliable only for $\omega$ not too large
\cite{gorsky_manohar}, e.g. $\vert\omega\vert < 1/2$. The case of one
on-shell photon corresponds to $\vert\omega\vert = 1$, so that the
coefficient of the $1/Q^2$ fall-off of the form factor
$\F_{\pi\gamma^*\gamma^*}(-Q^2,0)$ is actually not known.  Depending
on the assumptions made or the ans\"atze considered for
$\varphi_{\pi}(u)$, different results have been obtained in the
literature.

The LMD ansatz for $\H_V(q_1^2,q_2^2, (q_1+q_2)^2)$ reproduces these
results for $\vert\omega\vert < 1$ with
$f^{LMD}(\omega)=1/(1-\omega^2)$. On the other hand, taking $q^2_2=0$
and letting $Q^2=-q^2_1$ become large, we obtain
\bea
\F_{\pi\gamma^*\gamma^*}^{LMD}(-Q^2,0)\sim{\mbox{const}}.
\nonumber
\eea
In order to recover the $1/Q^2$ behaviour for $\vert\omega\vert = 1$,
we need to go beyond the LMD approximation and add a second vector
resonance (adding a pseudoscalar resonance \cite{Moussallam95} would
not help to improve the situation in the present case, since the
$1/Q^2$ behaviour forces $\hp_1 = 0$ in the ansatz for $\H_V$, in
contradiction with the relation (\ref{OPE_hP1})).  From
Eq.~(\ref{FGHVA_LMD+V}) we obtain
\bea
\F_{\pi\gamma^*\gamma^*}^{LMD+V}(q_1^2,q_2^2) =  
 {F_0 \over 3} { q_1^2 q_2^2 \left[
q_1^2 + q_2^2 \right] + \hv_1 (q_1^2 + q_2^2)^2 + \hv_2 q_1^2
q_2^2 + \hv_5 (q_1^2 + q_2^2) + \hv_7 \over (q_1^2 -
M_{V_1}^2) (q_1^2 - M_{V_2}^2) (q_2^2 - M_{V_1}^2) (q_2^2 -
M_{V_2}^2)} \, . \label{F_pigg_offshell_LMD+V}
\eea
The behaviour of $\F_{\pi\gamma^*\gamma^*}^{LMD+V}(q_1^2,q_2^2)$ for
$Q^2$ large and $\omega$ fixed, with $\vert\omega\vert <1$, is the same
as in the case of
$\F_{\pi\gamma^*\gamma^*}^{LMD}(q_1^2,q_2^2)$. However, if we now set
$q_2^2=0$, we obtain
\bea
\F_{\pi\gamma^*\gamma^*}^{LMD+V}(-Q^2,0)  = 
 {F_0 \over 3} {1 \over M_{V_1}^2 M_{V_2}^2} { \hv_1 Q^4 - \hv_5
Q^2 + \hv_7 \over (Q^2 + M_{V_1}^2) (Q^2 + M_{V_2}^2)} \, .
\eea
Imposing that this expression exhibits the  $1/Q^2$ behaviour for large 
$Q^2$ requires that $\hv_1$ vanishes, which then gives
\bea
\F_{\pi\gamma^*\gamma^*}^{LMD+V}(-Q^2,0)  = 
- \frac{2F_0}{3}\,\frac{1}{Q^2}\,\frac{\hv_5}{2M_{V_1}^2 M_{V_2}^2} + 
{\cal O}\left(\frac{1}{Q^4}\right) \, . 
\eea
In the absence of a reliable prediction for the coefficient that
governs the $1/Q^2$ behaviour, this still leaves the parameter $\hv_5$
undetermined.  Additional information may be obtained from the fact
that the form factor $\F_{\pi\gamma^*\gamma^*}(q_1^2,q_2^2)$ can also
be related to the decay $\rho^+\to\pi^+\gamma$, whose amplitude is
given by
\be
{\cal A}(\rho^+ \to \pi^+ \gamma)  =  \, {e \over 3} \, 
\lim_{q_1^2 \to M_V^2}\,\lim_{ q_2^2 \to 0}\, {(q_1^2 - M_V^2) \over F_V M_V}
\F_{\pi\gamma^*\gamma^*}(q_1^2,q_2^2) \, .  \label{amp_rhopigamma} 
\ee
Note that we obtain the same relation for ${\cal A}(\rho^0 \to \pi^0
\gamma)$. In the latter case, however, $\rho-\omega$ mixing would have 
to be taken into account for a realistic calculation.  Furthermore, we
have defined the coupling $F_V$ of the $\rho$ meson to the vector
current by
\bea
\langle 0 \vert V_\mu^a(0) \vert \rho^b(p) \rangle = \delta^{ab} F_V M_V
\varepsilon_\mu  \, ,  \nonumber 
\eea
where $\varepsilon_\mu$ denotes the polarization vector of the
$\rho$ meson. From Eq.~(\ref{amp_rhopigamma})
we obtain (see also \cite{Moussallam95})
\be
- \left( { 2 e F_V \over M_V} \right)  { {\cal A}(\rho^+ \to \pi^+
\gamma) \over {\cal A}(\pi^0 \to \gamma \gamma)}  =    
{ \lim_{q_1^2 \to M_V^2, q_2^2 \to 0} (q_1^2 - M_V^2)
\F_{\pi\gamma^*\gamma^*}(q_1^2,q_2^2)  \over 
\lim_{q_1^2 \to 0, q_2^2 \to 0} (q_1^2 - M_V^2)
\F_{\pi\gamma^*\gamma^*}(q_1^2,q_2^2) } =  1 + x \, .  \label{def_x}
\ee
The observed value $\Gamma = 68 \pm 7~\mbox{keV}$ for the decay width
$\rho^+ \to \pi^+ \gamma$~\cite{PDG2000} yields $x = 0.022 \pm 0.051$
\cite{Moussallam95}.  The LMD ansatz for $\F_{\pi\gamma^*\gamma^*}$
leads to $x_{LMD}=-(4\pi^2/N_C)(F_0^2/M_V^2)=-0.19$, far from the
experimental value~\cite{Moussallam_private_comm}.  Starting instead
from the form factor $\F_{\pi\gamma^*\gamma^*}^{LMD+V}$
leads to 
\bea
(1 + x)_{LMD+V} = {1 \over (1 - M_{V_1}^2 / M_{V_2}^2) }
\left( 1 - {4\pi^2 \over N_C} {F_0^2 \over M_{V_1}^2} \left[
{M_{V_1}^2 \over M_{V_2}^2} {\hv_1 \over M_{V_2}^2} + {\hv_5 \over
M_{V_2}^4} \right] \right) \, . \label{x_LMD+V_2} 
\nonumber
\eea
Setting $\hv_1=0$ and solving for $\hv_5$ gives $\hv_5=6.3\pm
0.9$~GeV$^4$, where we have taken $M_{V_1} = 769~\mbox{MeV}$, $M_{V_2}
= 1465~\mbox{MeV}$ for the resonance masses and $F_0 =
92.4~\mbox{MeV}$.

Actually, the form factor $\F_{\pi\gamma^*\gamma^*}(-Q^2,0)$ has been
measured in the space-like region by the CLEO collaboration, see Table
1 in Ref.~\cite{CLEO}, over a wide range of $Q^2$,
$1.5~\mbox{GeV}^2\le Q^2\le 9~\mbox{GeV}^2$. A fit of the expression
$\F_{\pi\gamma^*\gamma^*}^{LMD+V}(-Q^2,0)$ with $\hv_1 = 0$ to these
data yields
\bea \label{hV5_fit_2} 
\hv_5 = 6.93 \pm 0.26~\mbox{GeV}^4 \, , 
\eea
with $\chi^2/\mbox{dof} = 7.00/{14} = 0.50$. Keeping also $\hv_1$ as a free 
parameter yields instead
\bea
\hv_1 & = & -0.01 \pm 0.16~\mbox{GeV}^2 \, , \label{hV1_fit_1}\nonumber
 \\
\hv_5 & = & 6.88 \pm 0.61~\mbox{GeV}^4 \, ,  \label{hV5_fit_1}\nonumber
\eea
with $\chi^2/\mbox{dof} = 6.99 /{13} = 0.54$. The results for $\hv_5$
from both fits are compatible with the value extracted above from the 
decay $\rho^+ \to \pi^+ \gamma$. 

Finally, one might ask to which extent the inclusion of a second
vector resonance into the ansatz for $\H_V$ modifies the determination
of the corresponding combinations of low-energy constants. The only
combination which can be fixed from the knowledge of
$\F_{\pi\gamma^*\gamma^*}(-Q^2,0)$ alone is $A_{V,p^2}$ (see
Eq.~(\ref{A_combinations})),
\bea
A_{V,p^2}^{LMD+V}  =  {1\over 8} {F_0^2 \over M_{V_1}^4} {\hv_5
\over M_{V_2}^4} - {N_C \over 32 \pi^2} {1\over M_{V_1}^2} \left( 1 +
{M_{V_1}^2 \over M_{V_2}^2 } \right) \, ,  \label{AVp2_LMD+V}
\eea
since the other combination, $A_{V,(p+q)^2}^{LMD+V}$, involves
$\hv_6$. With the value of $\hv_5$ obtained in Eq.~(\ref{hV5_fit_2}),
we find, in units of $10^{-4}/F_0^2$, $A_{V,p^2}^{LMD+V}=-1.36$,
i.e. about 20~\% away from our LMD estimate reported in
Table~\ref{tab:A_i_LMD_res}. The difference is well within the 30\%
relative error that we attribute to the approximations considered
there.  

Another approach was followed in Ref.~\cite{Moussallam95}, where an
additional pseudoscalar resonance $\pi^\prime$ was included in the
ansatz for $\H_V$ that satisfies the OPE constraints (LMD+P, see
Eq.(\ref{H_V_LMD+P})).  As noted above, this ansatz will, however, not
correctly reproduce the $1/Q^2$ behaviour of
$\F_{\pi\gamma^*\gamma^*}(-Q^2,0)$ at large $Q^2$.  In this case,
$A_{V,p^2} = - N_C (1+x) / 32\pi^2 M_V^2$ and from Eq.~(\ref{def_x})
one obtains the result, again in units of $10^{-4}/F_0^2$,
$A_{V,p^2}^{LMD+P} = - 1.40$, close to our LMD+V estimate from the fit
to the CLEO data. In a similar way, the low-energy constant
$A_{V,(p+q)^2}$ receives within the LMD+P ansatz an additional
contribution proportional to the decay amplitude ${\cal A}(\pi^\prime
\to \gamma \gamma)$~\cite{Moussallam95}. Since this decay has not yet
been observed experimentally, our LMD estimate for $A_{V,(p+q)^2}$
from Table~\ref{tab:A_i_LMD_res} is probably not strongly modified by
the addition of a pseudoscalar resonance. 

The low-energy constant $A_{V,p^2}$ also describes the contributions
from the counterterms at order $p^6$ to the slope $b_\pi$ of the form
factor at the origin
\bea 
b_\pi \equiv \left( {1 \over {\cal A}(\pi^0 \to \gamma \gamma^*(q^2))}
{{\rm d} \over {\rm d} q^2} {\cal A}(\pi^0 \to \gamma \gamma^*(q^2))
\right)_{q^2 = 0} \, . \nonumber 
\eea 
At $\order(p^6)$ this slope also receives contributions from chiral
loops, therefore, $b_\pi = b_\pi^{loops} + b_\pi^{CT}$, with
$b_\pi^{CT} = - 32\pi^2 A_{V,p^2} / N_C$.  The Particle Data Group
gives the value $a_\pi \equiv M_{\pi^0}^2 b_\pi = 0.032 \pm
0.004$~\cite{PDG2000}.  From our LMD+V estimate above we obtain
$b_\pi^{CT, LMD+V} = 1.67~\mbox{GeV}^{-2}$ or, equivalently,
$a_\pi^{CT, LMD+V} = 0.031$. To this value, one should add the
contribution from the chiral logarithms, evaluated at the scale
$\mu\sim M_{\rho}$, $a_\pi^{loops}\sim 0.005$~\cite{bramon}, which
represents a 20\% effect. We note that the value for $a_\pi$ used by
the PDG is essentially the one reported by the CELLO
collaboration~\cite{CELLO}. In the latter paper a simple VMD-inspired
pole ansatz was fitted with their data for the form factor in the
space-like region for $0.5~\mbox{GeV}^2 \leq Q^2 \leq
2.7~\mbox{GeV}^2$, not taking into account contributions from
Goldstone boson loops at low $Q^2$.

\newpage
\section{Conclusions}

In this article we have studied the QCD three-point functions $\VAP,
\VVP$ and $\AAP$ in the three-flavour chiral limit in order to obtain
resonance estimates for some of the low-energy constants that appear
at ${\cal O}(p^6)$ in chiral perturbation theory in the meson sector
(even and odd intrinsic parity). We have compared the results that
have been obtained in the literature using a Lagrangian that includes
resonance fields \cite{Prades} with those evaluated within the
framework of an approximation of large-$N_C$ QCD combined with
information on short-distance properties.  In certain cases, we have
found substantially different results for the estimates of the
low-energy constants obtained with these two methods. We have pointed
out that this is due to the fact that the Green's functions derived
from the resonance Lagrangian do not correctly reproduce the QCD
short-distance behaviour. This defect can be repaired, but at the
expense of introducing, into the resonance Lagrangian, certain local
contributions. The difference with the similar situation at the ${\cal
O}(p^4)$ level lies in the fact that these local contributions {\it
cannot be restricted to terms already present} in the ${\cal O}(p^6)$
chiral Lagrangian, but also involve terms with resonance fields and
higher order derivatives. This feature, already noticed in a
particular case in \cite{Moussallam:1994at}, seems to be of a general
character. A general construction remains to be done, and appears to
be a much more complicated task than at $\order(p^4)$.

We note that although in general the short-distance behaviour of the
Green's functions derived from the resonance Lagrangian~\cite{Prades}
is incompatible with QCD, one can sometimes reproduce the results from
the operator product expansion, by adjusting the resonance parameters
accordingly.  There are, however, certain cases where this is not
possible.  The numerical values for the low-energy constants can then
be very different.  In particular, whereas both methods lead to
identical estimates for the resonance contributions at order $p^6$ in
the vector form factor of the pion $F_V^\pi(q^2)$, our estimate for the
resonance contribution in one of the form factors for the decay $\pi
\to e \nu_e \gamma$ is a factor of five smaller than the results
quoted in Ref.~\cite{BT}.

Of course one might argue that the short-distance behaviour of Green's
functions, i.e.\ their behaviour at very high energies, is irrelevant
for the determination of low-energy constants in chiral Lagrangians
starting from a resonance Lagrangian that is supposed to be valid only
in the intermediate energy region anyway. We think, however, that
taking into account the QCD short-distance constraints is a good
guiding principle to avoid (some of) the ambiguities when working with
resonance fields, as was shown at order $p^4$ in
Ref.~\cite{Ecker:1989yg}. Moreover, in certain cases one needs also
integrals of these Green's functions, for instance, to estimate the
low-energy constants that appear if virtual photons~\cite{Urech} or
leptons~\cite{Knecht:2000ag} are included in chiral perturbation
theory, see the discussion in Refs.~\cite{Moussallam97,KNinprep}. The
case of the counterterms of the effective
Lagrangians~\cite{nonleptonic} in the $|\Delta S| = 1$ or $|\Delta S|
= 2$ sectors of the Standard Model presents very similar
features~\cite{Qoperators,Q7_Q8}. In these applications it is crucial
that the Green's functions respect the short-distance constraints in
order to obtain ultraviolet finite results or to implement a correct
matching between long and short distances.

Finally, we wish to add a few remarks concerning the methodology
followed in the present study. We shall not come back on the use of
the large-$N_C$ framework, which seems to be unavoidable once
correlators of higher rank than two-point functions need to be
considered. Our approach deviates however from a full large-$N_C$
limit of QCD by at least two aspects. For one thing, we have only
imposed constraints coming from the leading (for the three-point
functions) and next-to-leading (for the vertex functions)
short-distance properties of QCD. We have considered neither the
effects of higher dimension operators in the Wilson expansions, nor
have we included QCD corrections to the short-distance behaviour
(recall however, that for the cases treated here, the corresponding
Wilson coefficients had no anomalous dimensions). For the other thing,
we have truncated the mesonic spectrum of large-$N_C$ QCD to the
minimal number of resonances necessary in order to fulfill the
short-distance constraints that were considered. Both approximations
are to a large extent interdependent. It is clear that, say, the
simplest LMD ansatz will at some point fail to reproduce the
subleading short-distance behaviour. On the other hand, the knowledge
of the subdominant operators in the OPE, or other constraints, will
fix additional parameters that have to be introduced if one goes
beyond the LMD approximation. In this sense, the framework within
which we have been working is, given the necessary amount of work,
improvable.  As an illustration, we have considered the
pion-photon-photon form factor $F_{\pi\gamma^*\gamma^*}(-Q^2,0)$.  The
experimentally observed~\cite{CLEO} $1/Q^2$ fall-off of the form
factor cannot be reproduced with our LMD ansatz for the invariant
function $\H_V$. Including one additional vector-resonance, we obtain
a representation for this form factor that fulfills all constraints
from the leading terms in the operator product expansion and that fits
the experimental results successfully if we adjust some of the unknown
parameters that enter in the generalized ansatz for $\H_V$. We note
that this is not possible if one includes a pseudoscalar resonance
instead. This new ansatz, with the phenomenologically determined
parameters, is furthermore compatible with the observed decay rate for
$\rho^+ \to \pi^+ \gamma$. Finally, the estimate for one combination
of low-energy constants $A_i$ in the odd-intrinsic parity sector
changes by about 20~\% with this new ansatz. This difference is within
the 30~\% relative error that we attribute to the approximations
considered here. The determination of the low-energy constants seems
thus to be stable against the inclusion of higher mass resonances.

It is clear that the three-point functions we have considered do not,
by far, exhaust the whole set of low-energy constants of the chiral
Lagrangian at the ${\cal O}(p^6)$ level. Other three point functions,
as well as higher correlators, need to be studied in a similar way for
that purpose. Some of them (in particular, those describing the
low-energy constants related to quark mass corrections) will include
the scalar densities. Recent studies have emphasized that in the
$0^{++}$ channel, predictions relying on the large-$N_C$ picture might not 
be very reliable~\cite{scalar,Q7_Q8}, because of 
the strong $\pi\pi$ interaction in the
S-wave. It has been suggested that a more appropriate
treatment would rather require to consider the limit where both $N_C$
and $N_F$, the number of light flavours, become large, with a fixed
ratio $N_F/N_C$ \cite{Veneziano:1976wm}.  Also, in the case of two-
and three-point functions, the quantum numbers of the resonances that
can contribute in the large-$N_C$ limit are entirely fixed by the
quantum numbers of the quark bilinears involved. This is no longer true
for $n$-point functions with $n\ge 4$~\cite{witten}. We leave these and other
interesting issues for future work.

\section*{Acknowledgements} 

We thank J.~Gasser, H.~Leutwyler, B.~Moussallam, M.~Perrottet and
J.~Stern for useful discussions. We are grateful to S.~Peris,
J.~Prades and E.~de Rafael for discussions and comments on the
manuscript. This work has been supported in part by Schweizerischer
Nationalfonds and by TMR, EC-Contract No.\ ERBFMRX-CT980169
(EURODA$\Phi$NE).


\appendix

\section{Operator product expansion for vertex functions}
\label{app:OPE}
\renewcommand{\theequation}{\Alph{section}.\arabic{equation}}
\setcounter{equation}{0}

In this appendix we discuss the short-distance behaviour of the vertex
functions
\bea
\left( \Gamma_{VA}\right)_{\mu\nu}^{abc}(q,p) &=&
\int d^4x e^{i q \cdot x} \langle 0 \vert T \{ V_\mu^a(x)
A_\nu^b(0) \} \vert \pi^c(p) \rangle \, , 
\nonumber\\
\nonumber\\
\left( \Gamma_{VP}\right)_{\mu}^{abc}(q,p) &=&
\int d^4x e^{i q \cdot x} \langle 0 \vert T \{ V_\mu^a(x)
P^b(0) \} \vert \pi^c(p) \rangle \, , 
\nonumber\\
\nonumber\\
\left( \Gamma_{VV}\right)_{\mu\nu}^{abc}(q,p) &=&
\int d^4x e^{i q \cdot x} \langle 0 \vert T \{ V_\mu^a(x)
V_\nu^b(0) \} \vert \pi^c(p) \rangle \, , 
\nonumber\\
\nonumber\\
\left( \Gamma_{AA}\right)_{\mu\nu}^{abc}(q,p) &=&
\int d^4x e^{i q \cdot x} \langle 0 \vert T \{ A_\mu^a(x)
A_\nu^b(0) \} \vert \pi^c(p) \rangle \, .   
\eea
They are related, according to the LSZ reduction formula, 
to the $\VAP$, $\VVP$ and $\AAP$ three-point functions 
as follows:
\bea
\left( \Gamma_{VA}\right)_{\mu\nu}^{abc}(q,p) & = & i {F_0 \over
\qbarq} f^{abc} \Bigg[ \qbarq \left( {(2p-q)_\mu (p-q)_\nu \over
(p-q)^2} - \eta_{\mu\nu} \right)  \nonumber\\
& & + P_{\mu\nu}(q,p-q) \tilde \F(q^2, q\cdot p) 
+ Q_{\mu\nu}(q,p-q) \tilde \G(q^2, q\cdot p) \Bigg] \, , 
\label{VA_explicit} \nonumber\\
\tilde \F(q^2, q\cdot p) & = & \lim_{p^2 \to 0} p^2 \F(q^2, (p-q)^2, p^2)
\, , \label{def_F_tilde} \nonumber\\
\tilde \G(q^2, q\cdot p) & = & \lim_{p^2 \to 0} p^2 \G(q^2, (p-q)^2, p^2)
\, , \label{def_G_tilde}
\eea
and
\bea
\left( \Gamma_{VP}\right)_{\mu}^{abc}(q,p) & = & {-1\over F_0}
f^{abc} \left[ - \qbarq {(q-2p)_\mu \over (q-p)^2} + 
(q^2 p_\mu - (q \cdot p) q_\mu) \check{\G}(q^2, q \cdot p) \right] \, , 
\label{VP_explicit} \nonumber\\
\check{\G}(q^2, q \cdot p) & = & \lim_{p^2 \to 0} p^2 \G(q^2, p^2,
(q-p)^2) \, ,  \label{def_G_check} 
\eea
since the invariant function $\F(q^2,p^2,(q-p)^2)$ does not contain a
$1/p^2$ Goldstone boson pole. 
Finally,
\bea
\left( \Gamma_{VV}\right)_{\mu\nu}^{abc}(q,p) & = & i\,{F_0 \over \qbarq}\, 
\epsilon_{\mu\nu\alpha\beta} q^\alpha p^\beta d^{abc} \, \tilde
\H_V(q^2, q\cdot p) \, ,  \nonumber\\ 
\left( \Gamma_{AA}\right)_{\mu\nu}^{abc}(q,p) & = & i\,{F_0 \over \qbarq}\, 
\epsilon_{\mu\nu\alpha\beta} q^\alpha p^\beta d^{abc} \, \tilde
\H_A(q^2, q\cdot p) \, ,  \nonumber\\  
\tilde \H_X(q^2, q\cdot p) & = & \lim_{p^2 \to 0} p^2 \H_X(q^2,
(p-q)^2, p^2) \, , \quad X = V,A \, , 
\eea
with 
$\tilde \H_X(q^2, q\cdot p) =  \tilde \H_X(q^2 - 2 q\cdot p, -q\cdot p)$,
as a consequence of Eq.~(\ref{Bose_symmetry}).

For general Dirac matrices $\Gamma_{1,2}$ we obtain the OPE
\bea
\lefteqn{ \lim_{\lambda \to \infty} \int d^4 x e^{i (\lambda q) \cdot x} 
\langle 0 \vert T \{ (\bar \psi \Gamma_1 {\lambda^a \over 2}  
\psi)(x) (\bar \psi  \Gamma_2 {\lambda^b \over 2} \psi)(0) \} 
\vert \pi^c(p) \rangle } \nonumber \\
& = & 
\lim_{\lambda \to \infty} \int d^4 x e^{i (\lambda q) \cdot x} 
(\Gamma_1)_\alpha^{~\beta} \left({\lambda^a \over
2}\right)_{IJ} (\Gamma_2)_\gamma^{~\delta} \left({\lambda^b \over
2}\right)_{KL} \nonumber \\ 
& & \qquad \times 
\Bigg[ i S(x)_\beta^{~\gamma} \delta_{JK} 
\langle 0 \vert : \bar \psi_I^\alpha(x) \psi_{\delta,L}(0) : 
\vert \pi^c(p) \rangle 
%
+ i S(-x)_\delta^{~\alpha} \delta_{IL} 
\langle 0 \vert : \bar \psi_K^\gamma(0) \psi_{\beta,J}(x) : 
\vert \pi^c(p) \rangle \Bigg] 
\nonumber\\
&& \qquad + \cdots \nonumber \\
& = &
\frac{i}{\lambda}
(\Gamma_1)_\alpha^{~\beta} \left({\lambda^a \over 2}\right)_{IJ} 
(\Gamma_2)_\gamma^{~\delta} \left({\lambda^b \over 2}\right)_{KL} 
  \nonumber \\
& & \quad \times \Bigg[ 
\left( {\qs \over q^2} \right)_\beta^{~~\gamma} \delta_{JK} 
\langle 0 \vert : \bar \psi_I^\alpha(0) \psi_{\delta,L}(0) : 
\vert \pi^c(p) \rangle 
- \left( {\qs \over q^2} \right)_\delta^{~~\alpha}
\delta_{IL}   
\langle 0 \vert : \bar \psi_K^\gamma(0) \psi_{\beta,J}(0) : 
\vert \pi^c(p) \rangle \Bigg]\nonumber \\
& &
+\frac{1}{\lambda^2}
(\Gamma_1)_\alpha^{~\beta} \left({\lambda^a \over 2}\right)_{IJ} 
(\Gamma_2)_\gamma^{~\delta} \left({\lambda^b \over 2}\right)_{KL} 
  \nonumber \\ 
& & \quad \times\frac{\partial}{\partial q^{\rho}}\Bigg[
 \left( {\qs \over q^2} \right)_\beta^{~~\gamma}
\delta_{JK} 
\langle 0 \vert : (D_\rho \bar \psi)_I^\alpha(0)
\psi_{\delta,L}(0) : \vert \pi^c(p) \rangle 
\nonumber \\
& &\qquad\qquad 
- \left( {\qs \over q^2} \right)_\delta^{~~\alpha}
\delta_{IL} 
\langle 0 \vert : \bar \psi_K^\gamma(0) (D_\rho \psi)_{\beta,J}(0) : 
\vert \pi^c(p) \rangle 
\Bigg] \nonumber \\
& &
+ {\cal O}\left(\frac{1}{\lambda^3}\right) , 
\label{OPE_JJpi_1} 
\eea
up to possible $\order(\alpha_s)$ corrections. Note that the indices
$I,J, \ldots$ label both flavour and colour and that in this Appendix
$(\lambda^a / 2)_{IJ}$ denote the Gell-Mann matrices in flavour space
and the unit matrix in colour space, with ${\rm tr}(\lambda^a
\lambda^b) = 2 N_C \delta^{ab}$.  From invariance
under parity, Lorentz, flavour $SU(3)_V$ and colour $SU(3)_C$
transformations the matrix elements involved in Eq.~(\ref{OPE_JJpi_1})
can be expressed as
\bea
\langle 0 \vert : \bar \psi_I^\alpha(0)
\psi_{\delta,L}(0) : \vert \pi^c(p) \rangle 
& = & \sum_d \left(
{\lambda^d \over 2} \right)_{LI} \Bigg[  
(i \gamma_5)_\delta^{~\alpha} \langle 0 \vert : (\bar \psi
i \gamma_5 {\lambda^d \over 2} \psi)(0) : \vert \pi^c(p) \rangle  
\left( {- 1 \over 2 N_C} \right) \nonumber \\
&& \quad \qquad \qquad 
+ (\gamma_\sigma \gamma_5)_\delta^{~\alpha} \langle 0 \vert : (\bar \psi
\gamma^\sigma \gamma_5 {\lambda^d \over 2} \psi)(0) : \vert \pi^c(p) \rangle  
\left( {- 1 \over 2 N_C} \right) 
\Bigg]  \, , \nonumber 
\eea
\newpage
\bea
\lefteqn{ \langle 0 \vert : (D_\rho\bar \psi)_I^\alpha(0) 
\psi_{\delta,L}(0) : \vert \pi^c(p) \rangle } \nonumber\\ 
& = & \sum_d \left( {\lambda^d \over 2}\right)_{LI} \Bigg[ 
(i \gamma_5)_\delta^{~\alpha} \langle 0 \vert : (D_\rho \bar \psi
i\gamma_5 {\lambda^d \over 2}   \psi)(0) : \vert \pi^c(p) \rangle 
\left( {- 1\over 2 N_C} \right) \nonumber \\
& & \quad \qquad \qquad + (\gamma_\sigma \gamma_5)_\delta^{~\alpha}
\langle 0 \vert : (D_\rho\bar \psi \gamma^\sigma \gamma_5 
{\lambda^d \over 2}  \psi)(0) : \vert \pi^c(p) \rangle 
\left( {- 1\over 2 N_C} \right) 
\nonumber \\ 
& & \quad \qquad \qquad + (\gamma_5
\sigma_{\rho\sigma})_\delta^{~\alpha} \langle 0 \vert : (D^\sigma \bar
\psi i \gamma_5 {\lambda^d \over 2}  \psi)(0) : \vert \pi^c(p) \rangle
\left( {- 1\over 6 N_C} \right) \Bigg] \, 
, \nonumber 
\eea
\bea
\lefteqn{ \langle 0 \vert : \bar \psi_K^\gamma(0)
(D_\rho \psi)_{\beta,J}(0) : \vert \pi^c(p) \rangle } \nonumber\\
& = & \sum_d \left( {\lambda^d \over 2}\right)_{JK} \Bigg[ 
(i \gamma_5)_\beta^{~\gamma} \langle 0 \vert : (\bar \psi
 i\gamma_5 {\lambda^d \over 2} D_\rho \psi)(0) : \vert \pi^c(p) \rangle 
\left( {-1 \over 2 N_C } \right) \nonumber \\
& & \quad \qquad \qquad + (\gamma_\sigma \gamma_5)_\beta^{~\gamma}
\langle 0 \vert : (\bar \psi
\gamma^\sigma \gamma_5 {\lambda^d \over 2} D_\rho \psi)(0) : \vert \pi^c(p)
\rangle \left( {- 1 \over 2 N_C } \right) \nonumber \\ 
& & \quad \qquad \qquad + (\gamma_5
\sigma_{\rho\sigma})_\beta^{~\gamma}  
\langle 0 \vert : (\bar \psi i \gamma_5 {\lambda^d \over 2}
D^\sigma\psi)(0) : \vert 
\pi^c(p) \rangle \left( {1 \over 6N_C } \right) \Bigg] \, . \nonumber 
\eea
In the above, we have also made use of the equations of motion in the
chiral limit, $\not \!\!D\psi=0$.

Invariance under space-time translations, parity and charge
conjugation further yields
\bea
\langle 0 \vert (D_\rho \bar \psi i \gamma_5 {\lambda^d \over 2} 
\psi)(0) \vert  \pi^c(p) \rangle 
 =   \langle 0 \vert (\bar \psi  i \gamma_5 {\lambda^d \over 2}
D_\rho \psi)(0) \vert \pi^c(p) \rangle 
= \,\frac{i}{2}\, p_\rho\,\frac{\qbarq}{F_0} \delta^{dc} \, ,
\nonumber 
\eea
\bea
\langle 0 \vert (D_\rho \bar \psi  
\gamma_{\sigma} \gamma_5 {\lambda^d \over 2} \psi)(0) \vert  \pi^c(p)
\rangle  =   \langle 0 \vert (\bar \psi  \gamma_{\sigma} \gamma_5
{\lambda^d \over 2} D_\rho \psi)(0) \vert \pi^c(p) \rangle 
= \,\frac{1}{2} p_\rho p_\sigma F_0 \delta^{dc} \, . 
\nonumber 
\eea

Using these expressions, we deduce from Eq.~(\ref{OPE_JJpi_1}) the
following short-distance behaviour of the vertex functions
\bea
\lim_{\lambda \to \infty}
\left( \Gamma_{VA}\right)_{\mu\nu}^{abc}(\lambda q,p) &=& 
\frac{i}{\lambda q^2}\,F_0 f^{abc}
\Bigg\{ (p\cdot q) \eta_{\mu\nu} -q_{\mu}p_{\nu}-q_{\nu}p_{\mu}
\nonumber\\
&&\quad\quad
+\,\frac{1}{\lambda q^2}\,[q^2p_{\mu}p_{\nu}+(p\cdot q)^2 \eta_{\mu\nu}-
(p\cdot q) (q_{\mu}p_{\nu}+q_{\nu}p_{\mu})] \Bigg\} 
+ {\cal O}\left(\frac{1}{\lambda^3}\right) , 
\nonumber\\
\lim_{\lambda \to \infty}
\left( \Gamma_{VP}\right)_{\mu}^{abc}(\lambda q,p) &=& 
\,\frac{1}{\lambda q^2}\,\frac{\qbarq}{F_0}\,f^{abc}
\left\{q_{\mu} +\,\frac{2}{3}\, {(p\cdot q) q_{\mu}-q^2p_{\mu} \over
\lambda q^2} \right\} + {\cal O}\left(\frac{1}{\lambda^3}\right) \, , 
\nonumber\\
\lim_{\lambda \to \infty}
\left( \Gamma_{VV}\right)_{\mu\nu}^{abc}(\lambda q,p) &=& 
- \frac{i}{\lambda q^2}\,F_0 d^{abc}\,
\epsilon_{\mu\nu\alpha\beta}q^\alpha p^\beta
\left\{1 + \frac{p\cdot q}{\lambda q^2} \right\} +
{\cal O}\left(\frac{1}{\lambda^3}\right) \, , 
\nonumber\\
\lim_{\lambda \to \infty}
\left( \Gamma_{AA}\right)_{\mu\nu}^{abc}(\lambda q,p) &=& 
- \frac{i}{\lambda q^2}\,F_0 d^{abc}\,
\epsilon_{\mu\nu\alpha\beta}q^\alpha p^\beta
\left\{1 + \frac{p\cdot q}{\lambda q^2} \right\} +
{\cal O}\left(\frac{1}{\lambda^3}\right) \, . 
\label{OPE_JJpi}
\eea
The result for $\Gamma_{VP}$ contradicts the one given in Eq.~(55) in
Ref.~\cite{Moussallam97} where no term of order $1/\lambda^2$
appears~\cite{Moussallam_private_comm2}. Note that this term is
transverse, in accordance with the chiral Ward identities from
Eq.~\rf{WardXXP}. On the other hand, our result for the short-distance
expansion of $\Gamma_{VA}$ agrees with the one given in Eq.~(58) in
Ref.~\cite{Moussallam97}.


\section{The resonance Lagrangian}
\label{app:Lag_res}
\renewcommand{\theequation}{\Alph{section}.\arabic{equation}}
\setcounter{equation}{0}

The resonance Lagrangian from Ref.~\cite{Prades}, which generalizes
the one already given in Ref.~\cite{Ecker:1989yg}, reads (omitting
terms including a scalar resonance) 
\bea
\lag_{res} & = & \lag_V + \lag_A + \lag_{VV}^{(2)} + \lag_{AA}^{(2)} + 
\lag_{VA}^{(2)} \, , \nonumber \\
\lag_V & = & - {1\over 4} \langle \hat V_{\mu\nu} \hat V^{\mu\nu} - 2
M_V^2 \hat V_\mu \hat V^\mu \rangle 
- {1 \over 2 \sqrt{2}} \left( f_V \langle \hat V_{\mu\nu}
f_{+}^{\mu\nu} \rangle + i g_V \langle \hat V_{\mu\nu} \left[ u^\mu ,
u^\nu \right] \rangle \right) \nonumber 
\\ 
& & + i \alpha_V \langle \hat V_\mu \left[ u_\nu ,
f_{-}^{\mu\nu}\right] \rangle 
+ \beta_V \langle \hat V_\mu \left[ u^\mu , \chi_{-} \right] \rangle
\nonumber \\
& & + i \theta_V \varepsilon_{\mu\nu\alpha\beta} \langle \hat V^\mu
u^\nu u^\alpha u^\beta \rangle 
+ h_V \varepsilon_{\mu\nu\alpha\beta} \langle \hat V^\mu \left\{
u^\nu , f_{+}^{\alpha\beta} \right\} \rangle \, , \nonumber \\ 
\lag_A & = & - {1\over 4} \langle \hat A_{\mu\nu} \hat A^{\mu\nu} - 2
M_A^2 \hat A_\mu \hat A^\mu \rangle 
- {1 \over 2 \sqrt{2}} f_A \langle \hat A_{\mu\nu}
f_{-}^{\mu\nu} \rangle 
+ i \alpha_A \langle \hat A_\mu \left[ u_\nu ,
f_{+}^{\mu\nu}\right] \rangle \nonumber \\ 
& & + \gamma_A^{(1)} \langle \hat A_\mu u_\nu u^\mu u^\nu \rangle 
+ \gamma_A^{(2)} \langle \hat A_\mu \left\{ u^\mu , u^\nu u_\nu
\right\} \rangle 
+ \gamma_A^{(3)} \langle \hat A_\mu u_\nu \rangle \langle u^\mu u^\nu
\rangle  
+ \gamma_A^{(4)} \langle \hat A_\mu u^\mu \rangle \langle u^\nu
u_\nu \rangle \nonumber \\ 
& & + h_A \varepsilon_{\mu\nu\alpha\beta} \langle \hat A^\mu \left\{
u^\nu , f_{-}^{\alpha\beta} \right\} \rangle \, , \nonumber \\ 
\lag_{VV}^{(2)} & = & {1\over 2} \delta_V^{(1)} \langle \hat V_\mu
\hat V^\mu u_\nu u^\nu \rangle 
+ {1\over 2} \delta_V^{(2)} \langle \hat V_\mu u_\nu \hat V^\mu u^\nu
\rangle  
+ {1\over 2} \delta_V^{(3)} \langle \hat V_\mu \hat V_\nu u^\mu u^\nu
\rangle 
+ {1\over 2} \delta_V^{(4)} \langle \hat V_\mu \hat V_\nu u^\nu u^\mu
\rangle \nonumber \\ 
& & + {1\over 2} \delta_V^{(5)} \langle \hat V_\mu u^\mu \hat V_\nu
u^\nu + \hat V_\mu u_\nu \hat V^\nu u^\mu \rangle 
+ {1\over 2} \kappa_V \langle \hat V_\mu \hat V^\mu \chi_{+} \rangle  
+ {1\over 2} i \phi_V \langle \hat V_\mu \left[ \hat V_\nu ,
f_{+}^{\mu\nu} \right] \rangle \nonumber \\
& & + {1\over 2} \sigma_V \varepsilon_{\mu\nu\alpha\beta} \langle
\hat V^\mu \left\{ u^\nu , \hat V^{\alpha\beta} \right\} \rangle \, , 
\nonumber \\ 
\lag_{AA}^{(2)} & = & {1\over 2} \delta_A^{(1)} \langle \hat A_\mu
\hat A^\mu u_\nu u^\nu \rangle 
+ {1\over 2} \delta_A^{(2)} \langle \hat A_\mu u_\nu \hat A^\mu u^\nu
\rangle  
+ {1\over 2} \delta_A^{(3)} \langle \hat A_\mu \hat A_\nu u^\mu u^\nu
\rangle 
+ {1\over 2} \delta_A^{(4)} \langle \hat A_\mu \hat A_\nu u^\nu u^\mu
\rangle \nonumber \\ 
& & + {1\over 2} \delta_A^{(5)} \langle \hat A_\mu u^\mu \hat A_\nu
u^\nu + \hat A_\mu u_\nu \hat A^\nu u^\mu \rangle 
+ {1\over 2} \kappa_A \langle \hat A_\mu \hat A^\mu \chi_{+} \rangle  
+ {1\over 2} i \phi_A \langle \hat A_\mu \left[ \hat A_\nu ,
f_{+}^{\mu\nu} \right] \rangle \nonumber \\
& & + {1\over 2} \sigma_A \varepsilon_{\mu\nu\alpha\beta} \langle
\hat A^\mu \left\{ u^\nu , \hat A^{\alpha\beta} \right\} \rangle \, , 
\nonumber \\
\lag_{VA}^{(2)} & = & i A^{(1)} \langle \hat V_\mu \left[ \hat A_\nu ,
f_{-}^{\mu\nu} \right] \rangle 
+ i A^{(2)} \langle \hat V_\mu \left[ u_\nu , \hat A^{\mu\nu} \right]
\rangle  
+ i A^{(3)} \langle \hat A_\mu \left[ u_\nu , \hat V^{\mu\nu} \right]
\rangle  
+ B \langle \hat V_\mu \left[ \hat A^\mu , \chi_{-} \right] \rangle
\nonumber \\
& & + H \varepsilon_{\mu\nu\alpha\beta} \langle \hat V^\mu \left\{
\hat A^\nu , f_{+} ^{\alpha\beta} \right\} \rangle 
+ i Z^{(1)} \varepsilon_{\mu\nu\alpha\beta} \langle u^\mu u^\nu
\left\{ \hat A^\alpha , \hat V^\beta \right\} \rangle   
+ i Z^{(2)} \varepsilon_{\mu\nu\alpha\beta} \langle u^\mu \hat A^\nu
u^\alpha \hat V^\beta \rangle \, , \nonumber \\ 
& & \label{lag_res} 
\eea
where the vector fields describing the vector and axial vector
resonances have been denoted by $\hat V_\mu$ and $\hat A_\mu$,
respectively. In Eq.~(\ref{lag_res}) we employed the usual
notations~\cite{Ecker:1989te,Ecker:1989yg}  
\bea
\hat R_\mu & = & {1\over \sqrt{2}} \sum_{a=1}^8 \hat R_\mu^a
\lambda^a, \quad \hat R = \hat V, \hat A \, , \nonumber \\ 
\hat R_{\mu\nu} & = & \nabla_\mu \hat R_\nu - \nabla_\nu \hat R_\mu \,
, \nonumber \\ 
\nabla_\mu \hat R & = & \partial_\mu \hat R + \left[ \Gamma_\mu , \hat 
R \right] \, , \nonumber \\
\Gamma_\mu & = & {1\over 2} \left\{ u^\dagger \left[ \partial_\mu - i
(v_\mu + a_\mu) \right] u + u \left[ \partial_\mu - i
(v_\mu - a_\mu) \right] u^\dagger \right\} \, , \nonumber \\ 
f_{\pm}^{\mu\nu} & = & u F_L^{\mu\nu} u^\dagger \pm u^\dagger
F_R^{\mu\nu} u \, , \nonumber \\
u_\mu & = & i \left\{ u^\dagger \left[ \partial_\mu - i
(v_\mu + a_\mu) \right] u - u \left[ \partial_\mu - i
(v_\mu - a_\mu) \right] u^\dagger \right\} \equiv i u^\dagger D_\mu
U u^\dagger = u_\mu^\dagger \, ,  
\nonumber \\
\chi_{\pm} & = & u^\dagger \chi u^\dagger \pm u \chi^\dagger u \, ,
\nonumber 
\eea
where the symbol $\langle M \rangle$ stands for the trace of the
$3\times 3$-matrix $M$. The field $u$ is the square root of the
Goldstone boson field, $U = u^2$, whereas $v_\mu, a_\mu$ and $\chi$ 
denote the external sources.

Note that in the case of bilinear interactions of $\hat V_\mu$ and
$\hat A_\mu$, only terms with at most one trace in flavour space are
included in $\lag_{res}$, which is compatible with the large-$N_C$
suppression of Zweig rule violating contributions. Furthermore, no
pseudoscalar resonances have been incorporated in the resonance
Lagrangian in Ref.~\cite{Prades}.  As stressed in
Ref.~\cite{Ecker:1989yg}, terms from ${\cal L}_2 + {\cal L}_4$
involving Goldstone bosons have to be added to the resonance
Lagrangian as well in order to correctly reproduce the QCD
short-distance behaviour of certain Green's functions.


\end{document}